\documentclass{aa}  
\usepackage{lineno}
\usepackage{graphicx}
\usepackage{subcaption}
\usepackage{soul}
\usepackage{txfonts}
\usepackage{natbib}
\usepackage{caption}
\usepackage{comment}
\usepackage{rotating}
\usepackage{pdflscape}
\usepackage{subcaption} 
\usepackage{placeins}
\usepackage[bookmarks=false, colorlinks=true, citecolor=blue, linkcolor=blue]{hyperref}
\usepackage{hyperref} 
\usepackage{flafter} 
\usepackage{float}
\usepackage[maxfloats=256]{morefloats}
\def\kms{\,km\,s$^{-1}$}

\begin{document} 

\title
{Physical conditions around high-mass young star-forming objects via simultaneous observations of excited OH and methanol masers}%\\
%or\\
%Gas properties in the environment of HMYSOs using the methanol and OH masers}

   \author{A. Kobak
         \inst{1}
         \href{https://orcid.org/0000-0002-1206-9887}{\includegraphics[scale=0.5]{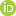}}
          \and
          A. Bartkiewicz
          \inst{1} \href{https://orcid.org/0000-0002-6466-117X}{\includegraphics[scale=0.5]{orcid.png}}
          \and
          K.~L.~J. Rygl
          \inst{2} 
          \href{https://orcid.org/0000-0003-4146-9043}{\includegraphics[scale=0.5]{orcid.png}}
          \and
          A. M. S. Richards
          \inst{3}
          \href{https://orcid.org/0000-0002-3880-2450}
          {\includegraphics[scale=0.5]{orcid.png}}
          \and
          M. Szymczak
          \inst{1}
          \href{https://orcid.org/0000-0002-1482-8189}
          {\includegraphics[scale=0.5]{orcid.png}}
          \and
          P. Wolak
          \inst{1}
          \href{https://orcid.org/0000-0002-5413-2573}
          {\includegraphics[scale=0.5]{orcid.png}}
         }
         
\institute{Institute of Astronomy, Faculty of Physics, Astronomy and Informatics, Nicolaus Copernicus University, Grudziadzka 5, 87-100 Torun, Poland
\and
INAF-Istituto di Radioastronomia, Via P. Gobetti 101, I-40129, Bologna, Italy
\and
JBCA, Department of Physics and Astronomy, University of Manchester, UK}

  \date{Received  2024 / Accepted 2024}

% \abstract{}{}{}{}{} 
% 5 {} token are mandatory
 
  \abstract
{Astrophysical masers are widely used in star formation studies. In particular, they are valuable in investigations of high-mass star-forming regions that are difficult to observe at optical frequencies.}
{We used multi-transition data to derive physical conditions in the immediate environment of forming high-mass stars.} 
{Simultaneous observations of two maser transitions, excited OH at 6.035\,GHz and methanol at 6.668\,GHz, were made using e-Merlin. Both transitions are radiatively pumped but prefer diverse physical conditions.}
{We imaged ten high-mass star-forming sites with milliarcsecond angular resolution, identifying regions where excited OH and methanol masers coexist and where they avoid each other. Moreover, we identified circularly polarized Zeeman splitting pairs of the OH transition, estimating magnetic field strengths in the range from 0.2 to 10.6~mG. The detection of linearly polarized components enabled us to compare the directions of magnetic field vectors with the outflows coming from the young star-forming objects.} 
{We found that the two maser lines appeared to coexist in six high-mass star-forming regions, in cloudlets separated by up to 205~au. Where the lines show avoidance, this can be related to changes in dust and gas temperatures; we also found a few examples suggestive of a high gas density. In seven sources, Kolmogorov-Smirnov tests show the nonrandom relationship between the position angles of distribution of the two maser transitions. We did not obtain consistent results regarding the direction of the magnetic field and outflow.}%; generally, the outflows tend to be perpendicular to the direction of linear polarization vectors.}

   \keywords{masers -- stars: massive -- stars: formation -- polarization -- ISM: magnetic fields -- ISM: molecules}

\titlerunning{Methanol and excited OH, coincidence and magnetic field}
\authorrunning{A. Kobak et al.}

   \maketitle

\section{Introduction}\label{sec:introd}

Astrophysical masers are used to study high-mass young star-forming objects (HMYSOs), in particular maser emission from hydroxyl (OH), water (H$_2$O), and methanol (CH$_3$OH) molecules. In this publication, we focus on two maser transitions: %that appear in protostellar discs are 
excited OH at 6.035\,GHz  (hereafter ex-OH) and methanol at 6.668\,GHz (hereafter 6.7\,GHz). Both are excited via thermal, infrared emission from warm dust that is heated by nearby protostars and may or may not occur in the same volume of gas, depending on the physical conditions, as was modeled in \cite{cragg2002}. 
The OH radical is a paramagnetic molecule; therefore, Zeeman splitting can be observed from the left- and right-hand circularly polarized (LHCP and RHCP) spectra, and the strength and direction of the magnetic fields can be derived from the separations of both features in the velocity domain (e.g., \citealt{avison2016}, \citealt{szymczak2020}, \citealt{ouyang2022}). If a third, linearly polarized component is detected at the central, unshifted velocity, a Zeeman triplet is identified (e.g., \citealt{hutawarakorn2002, green2015}).
As methanol is diamagnetic, the Zeeman splitting is much smaller and requires very high angular and velocity resolution for detection (e.g., \citealt{vlemmings2011}, \citealt{surcis2012}, \citealt{surcis2013}, \citealt{surcis2015}, \citealt{surcis2019}, \citealt{surcis2022}). 
Observations and modeling show that the 6.7~GHz transition is the brightest methanol maser under a wide range of physical conditions, such as a kinetic temperature of 30-150~K, appearing at dust temperatures above 100~K and diminishing when the gas temperature approaches or exceeds the dust temperature. The excitation of the 6.7 GHz methanol maser requires a hydrogen number density, $n$, in the range from 10$^5$ to 10$^{8.3}$\,cm$^{-3}$; the brightness temperature is independent of $n$ up to 10$^{8}$\,cm$^{-3}$ and above this value falls abruptly \citep{cragg2002}.
%Later, \cite{cragg2005} in their model extended this value even up to 10$^{9}$~cm$^{-3}$.  
The 6.035\,GHz OH emission arises from a main-line excited-state transition and appears at lower gas temperatures. It is quenched at a kinetic temperature above 70~K, and is independent of dust temperature. The ex-OH line is also related to high-number-density conditions, extending up to 10$^{8.5}$\,cm$^{-3}$ with a maximum brightness temperature in the range from 10$^7$ to 10$^{8}$\,cm$^{-3}$ \citep{cragg2002}.            
%The absence of the 6.035~GHz OH transition at low gas densities may indicate that collisions are involved in maser actions \citep{cragg2002}. 

The morphology and kinematics, including proper motion studies, of the methanol maser at 6.7\,GHz in HMYSOs have been relatively broadly examined (e.g., \citealt{bartkiewicz2009,bartkiewicz2014,bartkiewicz2016,bartkiewicz2020,bartkiewicz2024}, \citealt{goddi2011}, \citealt{moscadelli2011}, \citealt{pandian2011}, \citealt{sanna2010a,sanna2010b}, \citealt{sugiyama2008}). Still, it is not clear where exactly 6.7\,GHz methanol masers arise. In a few cases, the 6.7\,GHz maser reveals a ring-like morphology \citep{bartkiewicz2009}, which may indicate a disk seen face-on or an outflow oriented along the line of sight \citep{bartkiewicz2005}. However, detailed studies have shown that complex scenarios are common, in which the masers kinematics can result from a combination of rotation and expansion \citep{bartkiewicz2024}. \cite{sugiyama2011} measured the proper motions of 6.7\,GHz maser spots around the well-known HMYSO ON1 and concluded that the internal motions either traced the expanding ultracompact H{\sc ii} region or were associated with a molecular outflow. In HMYSO G23.01$-$0.41, the velocity field of methanol masers was explained in terms of a combination of slow radial expansion with rotation about an axis approximately parallel to the maser jet \citep{sanna2010b}.

In the case of ex-OH masers, far fewer regions have been imaged in this transition, and they tend to be more compact relative to the methanol masers, preventing a detailed discussion of the kinematic structures in general (e.g., \citealt{desmurs1998a,desmurs1998b,etoka2012}). The first untargeted survey of ex-OH, covering the Galactic longitude ranges 186\degr$<$l$<$60\degr, including the Galactic center, reported the detection of 127 sources, which is around~10\% of the number of sources with 6.7~GHz methanol masers found in this region \citep{avison2016}. In general, ex-OH masers are weaker than methanol masers, with brightness temperatures as much as two orders of magnitude lower. This statistic was confirmed by \cite{szymczak2020}, when they surveyed a sample of 445 HMYSOs, identified by the 6.7~GHz methanol transition, and detected 37 ex-OH targets, using the Torun 32-m radio telescope. Due to the small number of imaged regions, it is still uncertain where the ex-OH masers arise. They can be related to a disk, an outflow as well as an intermediate regions. Ex-OH masers may occur in the same place as ground-state (1.665 and 1.667\,GHz) OH masers. The latter can trace a dense molecular disk \citep{hutawarakorn2002}. However, \cite{caswell2011g24} present a case in which the ground-state OH emission arises in a fast-moving outflow.

Multi-transition studies of maser emission are valuable in constraining the physical conditions in the gas around HMYSOs. \cite{fish2007} used the European VLBI Network and presented the distributions of ex-OH in the HMYSO W3(OH) compared with ground-state OH masers. 
The ex-OH masers trace the inner edge of a counterclockwise rotating torus of dense molecular gas, while the ground-state OH occurs predominantly to the west of this edge. The 1.667\,GHz transition is generally associated with the ex-OH transition, but the 6.0\,GHz masers may also appear without the presence of the 1.667\,GHz masers. In contrast, 1.665\,MHz transition is ubiquitous and can appear separately as well as in association with the ex-OH transition.
Identified Zeeman pairs provided estimates of the magnetic field strength of a few milligauss and revealed a reversal of the line-of-sight direction of the magnetic field, oriented away from the observer in the west and toward the observer in the east.
\cite{green2007} imaged the 6.7\,GHz methanol and ex-OH lines toward the HMYSO ON1 using MERLIN. Toward the southern part of the ultracompact (UC)\,H{\sc ii} region, the two masers were observed to present a spatial distribution elongated perpendicular to the large-scale bipolar outflow. 
The Zeeman pairs identified in ex-OH implied the existence of a magnetic field oriented toward the observer with a strength of up to 5.8\,mG. The first tentative detection of Zeeman splitting in the 6.7~GHz methanol line was roughly consistent with the above value being $-$18$\pm$6\,mG (a negative value corresponds to the direction pointing toward the observer). Global VLBI observations toward the same target revealed that ex-OH masers have a distribution similar to that of the ground-state OH masers, in contrast to the situation in W3(OH), and this suggests the absence of the highest-density knots in ON1 \citep{fish2010}.  

This project aims to derive the physical conditions in a sample of HMYSOs via combined observations of ex-OH and methanol masers. In fact, the spatial coincidence or avoidance of these two masers implies specific physical conditions following the maser excitation models of \citet{cragg2002}. Moreover, the ex-OH maser observations provide: 1) the identification of Zeeman splitting, and thus estimates of the magnetic field strength, and 2) the detection of linearly polarized emission and estimation of the orientation of the magnetic field. This allows us to map the magnetic field structure in the close surroundings of HMYSOs. The continuum emission and methanol polarization will be reported in a future article.          

\section{Observations, data reduction, and methods}\label{sect:obser}

The source sample has been chosen based on a large survey of ex-OH carried out since 2018 with the Torun 32\,m telescope \citep{szymczak2020}. 
Among the 37 detections, we selected 10 sources in which the ex-OH line has a relatively strong (above 1.7~Jy) and non-variable flux density, as was found during two epochs of the survey (November-December 2018 and March-April 2019). In addition, our single-dish studies showed that the spectral features of the 6.7\,GHz methanol and ex-OH maser transitions appear at similar LSR velocities. %Ten of these 14 sources were observed successfully.

The observations were made using the enhanced Multi-Element Remotely Linked Interferometer Network (e-MERLIN)\footnote{e-MERLIN is a National Facility operated by the University of Manchester at Jodrell Bank Observatory on behalf of STFC. https://www.e-merlin.ac.uk/}, an array of six radio telescopes with baselines between 10--217\,km \citep{merlin2016}, under the project CY10206 in July--October 2020 and August 2022.
All observations were recorded in full polarization (RR, RL, LR, LL, where R and L represent right- and left-handed circular polarization), and averaged to 4-sec integrations. We observed the masers in two separate sets of spectral tunings, one set covering the ex-OH line (6035.092\,MHz) and the other the methanol line (6668.519\,MHz). 
In each tuning, the line was covered by a narrow spectral window 1\,MHz wide, consisting of 512 channels with a separation of 0.1\kms. Each observation was made at constant frequency. Four wide-band ($\sim$0.13\,GHz each) spectral windows were also observed in the frequency ranges of 6.00--6.51\,GHz and 6.30--6.81\,GHz (for the ex-OH and methanol tunings, respectively). 
For each individual source, the ex-OH and methanol tunings were interleaved, changing frequencies every few hours to ensure as near-simultaneous observations as was practical. 
In some cases, diﬀerent targets were also interleaved to optimize visibility plane coverage. The list of sources with their phase calibrators, total on-source time, and the central local standard of rest (LSR) velocities is given in Table~\ref{tab:config}. J1331$+$3030 (3C286) was observed as a flux and polarization angle standard. J0319$+$4130 (3C84) and J1407$+$2827 (OQ208) were observed as bandpass calibrators.

\begin{table*}
\centering
\caption{Details of eMERLIN observations.}
\begin{tabular}{cccccccc}
\hline
 \multicolumn{1}{c}{Source} & \multicolumn{1}{c}{Phase Calibrator} & \multicolumn{2}{c}{Number of scans} & \multicolumn{2}{c}{Total time on source} & \multicolumn{2}{c}{Center Vlsr} \\
 \multicolumn{1}{c}{name (l b)} & \multicolumn{1}{c}{} & \multicolumn{1}{c}{exOH} & \multicolumn{1}{c}{meth} & \multicolumn{1}{c}{exOH} & \multicolumn{1}{c}{meth} & \multicolumn{1}{c}{exOH} & \multicolumn{1}{c}{meth} \\
  (\degr\hspace*{0.5cm}\degr)       &                  &         &        &    (h:m)       &       (h:m)            &   (\kms)    &      (\kms)  \\

\hline
G20.237$+$0.065         &        J1818$-$1108   &       8       &       8       &       00:56   &       00:56   &       72.3    &       64.2 \\
G24.148$-$0.009         &        J1825$-$0737   &       9       &       6       &       01:03   &       00:42   &       16.7    &       9.5     \\
G25.648$+$1.049         &        J1827$-$0405   &       8       &       8       &       00:56   &       00:56   &       37.4    &       29.6    \\
%4      &        G28.201        &               &               &               &               &               &               &               \\
G34.267$-$0.210         &        J1858$+$0313   &       33      &       33      &       03:18   &       03:18   &       58.0    &       49.8    \\
G43.149$+$0.013         &        J1922$+$0841   &       33      &       33      &       03:18   &       03:18   &       13.5    &       5.2     \\
%7      &        G45.071$+$0.132        &               &               &               &               &               &               &               \\
%8      &        G45.445        &               &               &               &               &               &               &               \\
%9      &        G45.467        &               &               &               &               &               &               &               \\
G48.990$-$0.299         &        J1922$+$1530   &       10      &       10      &       01:10   &       01:10   &       56.4    &       48.3    \\
G49.490$-$0.388         &        J1924$+$1540   &       10      &       10      &       01:10   &       01:10   &       43.7    &       36.8    \\
G69.540$-$0.976         &        J2010$+$3322   &       12      &       12      &       01:24   &       01:24   &       $-$8.7  &       $-$16.4 \\
%13     &        G81.871$+$0.781(Jul)   &               &               &               &               &               &               &               \\
G81.871$+$0.781         &        J2018$+$3851   &       18      &       18      &       02:06   &       02:06   &       11.9    &       3.7     \\
G108.766$-$0.986        &        J2301$+$5706   &       13      &       12      &       01:31   &       01:24   &       $-$66.0 &       $-$73.9 \\
\hline
\end{tabular}
%Imstat dla kanału bez emisji (chan=15).
\label{tab:config}
\end{table*}

The data were processed using the e-MERLIN CASA 5.8 pipeline \citep{casadata}. The e-MERLIN pipeline includes time- and frequency-dependent calibration of the wide-band data, including estimating the flux scale in janskys and deriving corrections from the phase calibrator source.
We calibrated the narrow-band data using CASA 5.8 \citep{mcmullin2007}. The first step was flagging any remaining bad data. Then, we used the bandpass calibrator 3C84 to calculate corrections for the phase offset between the wide- and narrow-band data and applied these to the narrow-band 3C84 data, along with the pipeline time-dependent wide-band delay, phase, and amplitude corrections. Next, we performed frequency-dependent calibration to derive a bandpass correction table for amplitude and phase.
For ex-OH observations, we also derived corrections for polarization leakage (using 3C84, assumed to be unpolarized) and polarization angle by setting the Stokes parameter polarization flux densities for 3C286 for the known position angle (PA) of 33\degr~\citep{perley2013}.

Next, we applied the narrow-band bandpass calibration, the offset correction, the phase calibrator solutions, and, when applicable, the polarization calibration to the narrow-band target data. We split out the corrected target data, adjusting to constant velocity (LSR, radio convention, relative to the rest frequency of the line) in the target frame. 
Lastly, we self-calibrated each target using a selected bright channel; for ex-OH, we took care to preserve the relative flux densities in RR and LL, and thus the accuracy of Stokes $V$ (\cite{green2007, Darwish2020}, as implemented in CASA). We applied the solutions to all narrow channels and made spectral image cubes for each target (in full polarization, for ex-OH). 
Using uniform weighting to optimize resolution, the nominal 2-D Gaussian fits to the dirty beam varied from 121$\times$26\,mas at the lowest declination to 54$\times$20\,mas for the higher-declination targets (the long axis being close to N-S for the low-declination sources). Since the e-MERLIN beam is quite irregular, we used an intermediate weighting (robust 0.5) and a standard restoring beam of 65~mas$\times$45~mas at a PA of $-$50\degr, which has a slightly larger area than the nominal beam fits. This minimizes sidelobe artifacts and makes it easier to compare the images. Since we use component fitting to make measurements, the exact beam shape does not affect the results.

%\section{Methods}\label{sec:methods}
We used the JMFIT procedure in AIPS \citep[NRAO 2022;][]{greisen2003} to fit 2D Gaussian components to Stokes {\it I} (total intensity) maser spots for both the methanol and ex-OH masers, and thus measured the positions and the peak flux densities. 
In addition, using the positions of the I Stokes ex-OH maser spots, we measured their {\it Q} and {\it U} Stokes parameters as well as the LHCP and RHCP. For further analysis, we took only those spots with a total intensity above 2$\sigma$ that appear in at least two consecutive channels. Table \ref{tab:noise} lists the root mean square (rms) for all cubes measured in channels without emission.

\begin{table*}
\centering
\caption{Rms noises (1$\sigma$) in channels without emission for image cubes  in each Stokes parameter.} %in quiet channels of the datacubes for each source.}
\begin{tabular}{ccccccc}
\hline
 Source & $\sigma_{\mathrm{I\_meth}}$ & $\sigma_{\mathrm{I\_exOH}}$ & $\sigma_{\mathrm{Q\_exOH}}$ & $\sigma_{\mathrm{U\_exOH}}$ & $\sigma_{\mathrm{LHCP\_exOH}}$ & $\sigma_{\mathrm{RHCP\_exOH}}$  \\
name (l b)  & (mJy beam$^{-1}$) & (mJy beam$^{-1}$) & (mJy beam$^{-1}$) & (mJy beam$^{-1}$) & (mJy beam$^{-1}$) & (mJy beam$^{-1}$) \\
\hline
 G20.237$+$0.065 & 18.4 & 17.0 & 16.6 & 17.6 & 19.4 & 22.8 \\ 
 G24.148$-$0.009 & 28.7 & 12.7 & 12.3 & 12.6 & 16.9 & 15.9 \\
 G25.648$+$1.049 & 15.9 & 12.0 & 11.5 & 11.8 & 18.2 & 16.4 \\
% G28.201 & - & - & - & - & - & - \\
 G34.267$-$0.210 & 16.3 & 8.0 & 8.1 & 8.9 & 11.4 & 10.1 \\
 G43.149$+$0.013 & 14.3 & 15.2 & 16.5 & 16.0 & 22.3 & 20.6 \\
% G45.071$+$0.132 & 10.0 & - & - & - & - & - \\
% G45.445 & \\
% G45.467 & \\
 G48.990$-$0.299 & 55.6 & 15.9 & 16.4 & 15.9 & 23.0 & 20.7 \\
 G49.490$-$0.388 & 59.8 & 19.9 & 21.7 & 22.1 & 29.8 & 27.1 \\
 G69.540$-$0.976 & 18.3 & 17.8 & 18.2 & 17.5 & 20.6 & 21.1 \\
% G81.871$+$0.781(Jul) & 15.2 & 16.9 & 17.1 & 16.5 & 25.3 & 20.4 \\
 G81.871$+$0.781 & 8.5 & 9.9 & 9.8 & 10.2 & 14.9 & 13.9 \\
 G108.766$-$0.986 & 13.9 & 27.2 & 22.5 & 21.9 & 33.5 & 31.7 \\
\hline
\end{tabular}
%Imstat dla kanału bez emisji (chan=15).
\label{tab:noise}
\end{table*}

Tables \ref{tab:meth} and \ref{tab:exoh} summarize the properties of the brightest maser spot in each source: the absolute coordinates (RA, Dec), the LSR velocity ($V_{\mathrm{p}}$), and the peak flux density ($S_{\mathrm{p}}$). The astrometric accuracies presented in both tables were calculated following the procedure described in \cite{Richards2022}: 1) the uncertainty ($\delta_1$) of the position of the phase calibrator, taken from VLBA Catalog\footnote{https://obs.vlba.nrao.edu/cst/}, 2) the positional errors ($\delta_2$) due to the image rms noise, calculated for the weakest spots to get their upper limits, 3) phase errors ($\delta_3$) due to the phase-calibrator and target separation (including the time difference and the angular separation); values of $\delta_3$ range from 12 to 81\,mas, due to variations in atmospheric conditions, and 4) the antenna position errors ($\delta_4$) -- we assume a typical baseline error of 2~cm. The total astrometric accuracy is $\delta_{\mathrm{tot}}=(\delta_1^2+\delta_2^2+\delta_3^2+\delta_4^2)^{0.5}$. 

\begin{figure}[h!]
    \centering
    \includegraphics[width=0.85\columnwidth]{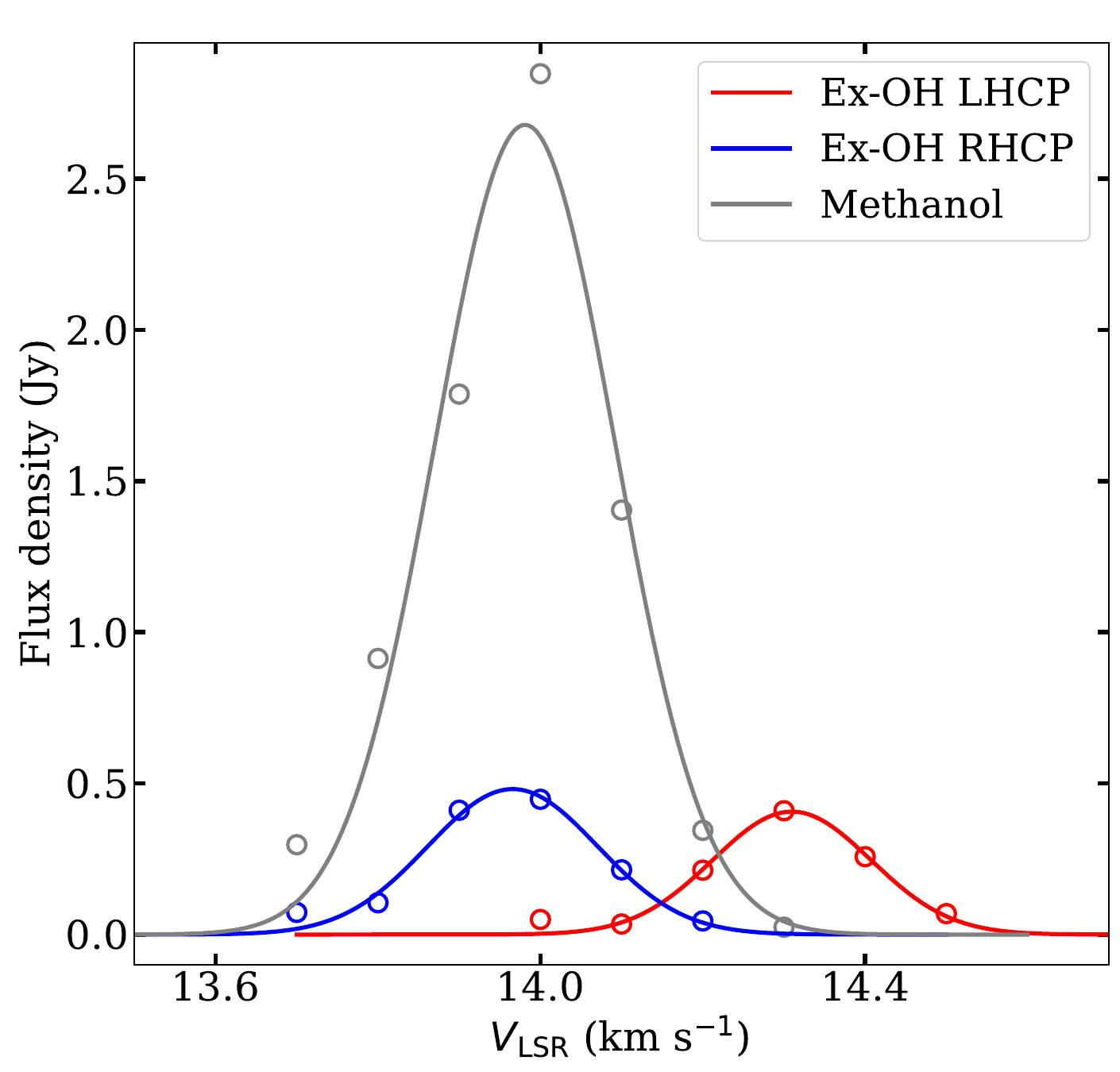}
    \caption{Example of the overlap of both maser transitions when Zeeman splitting is identified in the ex-OH transition. Profiles are from a maser group in G43.149$+$0.013 named as Z$_8$ in ex-OH (Table\,\href{https://doi.org/10.5281/zenodo.14865259}{C.1}). The dots correspond to the flux density measured on single-channel maps, and the lines are Gaussian profiles fit to the spectral features.}
    \label{fig:example_ZP}
\end{figure}

For the analysis of the coexistence of both maser transitions, we presumed that the lines coincide when: i) Their spectral features at least partially overlie each other in the LSR velocity range; we analyzed both polarizations of ex-OH to take into account the Zeeman splittings on this emission (see an example in Fig.~\ref{fig:example_ZP}). In the case of methanol emission, we neglected the splitting of the two polarizations due to the magnetic field, since its value is significantly smaller (less than one tenth) than the width of one spectral channel. ii) The angular separation between methanol and ex-OH spots is less than $0.5(\gamma_{6.7}+\gamma_{6.035})$, where $\gamma$ is calculated as ($\delta_2^2+\delta_3^2$)$^{0.5}$ for both transitions, respectively. Since we compare observations made with the same phase calibrator and antennas, we do not need to consider errors $\delta_1$ and $\delta_4$. The values of $\gamma$ are listed in Tables\,\ref{tab:meth} and\,\ref{tab:exoh}, respectively.

\begin{table*}
\centering
\caption{Parameters of the brightest methanol maser spot in each target.}
\begin{tabular}{ccccccc}
\hline
 Source & RA$_{S_{\mathrm{p}}^{6.7}}\pm\delta_{\mathrm{tot,RA}}$ & Dec$_{S_{\mathrm{p}}^{6.7}}\pm\delta_{\mathrm{tot,Dec}}$ & $V_{\mathrm{p}}^{6.7}$ & $S_{\mathrm{p}}^{6.7}$ & $T_{\mathrm{B}}$ & $\gamma_{6.7}$ \\
 name (l b) & (h~m~s\hspace{0.3cm}s) & (\degr~\arcmin~\arcsec\hspace{0.3cm}\arcsec) & (\kms)  & (Jy beam$^{-1}$) & $\times 10^9$(K) & (mas)\\ 
\hline
 G20.237$+$0.065 & 18 27 44.5613$\pm$0.0029 & $-$11 14 54.216$\pm$0.043 & 72.0 & 22.65 & 1.58 & 41.8 \\ 
 G24.148$-$0.009 & 18 35 20.9399$\pm$0.0014 & $-$07 48 55.775$\pm$0.021 & 17.6 & 15.32 & 1.12 & 19.0 \\
 G25.648$+$1.049 & 18 34 20.9069$\pm$0.0032 & $-$05 59 42.261$\pm$0.048 & 41.8 & 95.01 & 6.77 & 47.1 \\
%4 & G28.201 & - & - & - \\
 G34.267$-$0.210 & 18 54 37.2500$\pm$0.0029 & 01 05 33.615$\pm$0.043 & 54.5 & 5.90 & 0.41 & 42.2 \\
 G43.149$+$0.013 & 19 10 11.0464$\pm$0.0042 & 09 05 20.371$\pm$0.063 & 13.2 & 7.51 & 0.59 & 62.3 \\
%7 & G45.071$+$0.132 & 19 13 22.1263444 & 10 50 53.048216 & 31.52 \\
%8 & G45.445\\
%9 & G45.467\\
 G48.990$-$0.299 & 19 22 26.1312$\pm$0.0014 & 14 06 39.696$\pm$0.021 & 71.5 & 2.78 & 0.19 & 20.5 \\
 G49.490$-$0.388 & 19 23 43.9461$\pm$0.0018 & 14 30 34.366$\pm$0.027 & 59.2 & 613.87 & 41.44 & 25.8 \\
 G69.540$-$0.976 & 20 10 09.0436$\pm$0.0025 & 31 31 34.989$\pm$0.038 & 14.6 & 65.14 & 4.39 & 36.7 \\
%13a & G81.871$+$0.781(Jul) & 20 38 36.4053217 & 42 37 35.123346 & 249.05 \\
 G81.871$+$0.781 & 20 38 36.4097$\pm$0.0038 & 42 37 35.110$\pm$0.056 & 4.6 & 195.65 & 14.32 & 53.0 \\
 G108.766$-$0.986 & 22 58 51.1857$\pm$0.0018 & 58 45 14.380$\pm$0.026 & $-$45.7 & 12.25 & 0.83 & 25.4\\
\hline
\end{tabular}
\tablefoot{Note that the $\gamma$ parameter is estimated for the weakest methanol spot.}
\label{tab:meth}
\end{table*}

\begin{table*}
\centering
\caption{The parameters of the brightest ex-OH maser spots (Stokes I) in each target.} 
\begin{tabular}{ccccccc}
\hline
 Source & RA$_{S_{\mathrm{p}}^{6.035}}\pm\delta_{\mathrm{tot,RA}}$ & Dec$_{S_{\mathrm{p}}^{6.035}}\pm\delta_{\mathrm{tot,Dec}}$ & $V_{\mathrm{p}}^{6.035}$ & $S_{\mathrm{p}}^{6.035}$ & $T_{\mathrm{B}}$ & $\gamma_{6.035}$  \\
      name (l b)   &  (h~m~s\hspace{0.3cm}s) & (\degr~\arcmin~\arcsec\hspace{0.3cm}\arcsec) & (\kms)  & (Jy beam$^{-1}$) & $\times 10^7$(K) & (mas) \\ 
\hline
 G20.237$-$0.065 & 18 27 44.5623$\pm$0.0023 & $-$11 14 54.065$\pm$0.034 & 71.4 & 1.29 & 9.89 & 32.6\\
 G24.148$-$0.009 & 18 35 20.9398$\pm$0.0025 & $-$07 48 55.773$\pm$0.037 & 17.2 & 1.14 & 8.68 & 35.8 \\
 G25.648$+$1.049 & 18 34 20.9313$\pm$0.0034 & $-$05 59 42.655$\pm$0.051 & 39.5 & 1.55 & 12.72 & 49.7\\
%4 & G28.201 & - & - & - \\
 G34.267$-$0.210 & 18 54 37.2497$\pm$0.0030 & 01 05 33.622$\pm$0.045 & 54.1 & 1.91 & 14.62 & 44.1 \\
 G43.149$+$0.013 & 19 10 11.0632$\pm$0.0037 & 09 05 20.250$\pm$0.056 & 11.1 & 1.44 & 10.24 & 54.4\\
%7 & G45.071$+$0.132 & - & - & - \\
%8 & G45.445\\
%9 & G45.467\\
 G48.990$-$0.299 & 19 22 26.1350$\pm$0.0019 & 14 06 39.704$\pm$0.018 & 68.1 & 1.12 & 20.43 & 17.3 \\
 G49.490$-$0.388 & 19 23 43.9017$\pm$0.0015 & 14 30 33.548$\pm$0.023 & 54.8 & 5.32 & 40.14 & 22.1 \\
 G69.540$-$0.976 & 20 10 09.0859$\pm$0.0023 & 31 31 34.840$\pm$0.035 & 14.4 & 4.46 & 34.41 & 33.6 \\
%13a & G81.871$+$0.781(Jul) & 20 38 36.4237365 & 42 37 34.802839 & 2.80  \\
 G81.871$+$0.781 & 20 38 36.4209$\pm$0.0054 & 42 37 34.741$\pm$0.081 & 6.7 & 2.72 & 20.67 & 79.1\\
 G108.766$-$0.986 & 22 58 51.1786$\pm$0.0021 & 58 45 14.317$\pm$0.031 & $-$45.9 &  0.83 & 6.27 & 30.2 \\
\hline
\end{tabular}
\tablefoot{Note that the $\gamma$ parameter is estimated for the weakest ex-OH spot.}
\label{tab:exoh}
\end{table*}

\section{Results}\label{sec:results}
The distributions of the Stokes {\it I} maser spots and spectra of methanol and ex-OH are presented in the top panels on Figs.\,\ref{fig:g20p237}, \ref{fig:g24p148}, and \href{https://doi.org/10.5281/zenodo.14865259}{B.1}-\href{https://doi.org/10.5281/zenodo.14865259}{B.8}. When the coincidence of both transitions occurs, we mark them with non-filled symbols. The kinematic distances used to estimate linear sizes of sources were calculated using the parallax-based distance calculator\footnote{http://bessel.vlbi-astrometry.org/node/378} by \citet{reid2019}. In some cases, the ambiguity was resolved using the H{\sc i} absorption by \cite{green2011}. 
In the case of G48.990$-$0.299, G69.540$-$0.976, and G81.871$+$0.781, direct distance measurements were available via the trigonometric parallax method.  

The LHCP and RHCP ex-OH emission is presented in the middle panels of Figs:\,\ref{fig:g20p237}, \ref{fig:g24p148}, and \href{https://doi.org/10.5281/zenodo.14865259}{B.1}-\href{https://doi.org/10.5281/zenodo.14865259}{B.8}, where we also give the magnetic field strength along the line of sight derived from the identified Zeeman pairs. The relationship between Stokes {\it I} and both circular polarizations is: 0.5$\times$(RHCP$+$LHCP). As is noted in Sect.~\ref{sect:obser}, we measured the properties of RHCP and LHCP spots in the same position as I spots in each channel to ensure that we sampled the same gas when identifying Zeeman pairs. Only position error $\delta_2$ is applicable to comparisons within the same data cube. The worst accuracy for the faintest spots is $\sim$16\,mas, which is less than the physical size of the emitting regions, estimated from the angular extent of series of spots in consecutive channels, ensuring that detected associations between the polarizations are genuine.  

Table\,\href{https://doi.org/10.5281/zenodo.14865259}{C.1} lists the parameters of the identified Zeeman pairs: the flux density of the brightest maser spot in each polarization ($S_{\mathrm{max}}$) and the parameters of the fit Gaussian profile to the spectrum; the peak amplitude ($S_{\mathrm {fit}}$), the full width at half maximum of the profile ($FWHM_{\mathrm {fit}}$), and the peak velocity ($V_{\mathrm {fit}}$). 
We also list $\Delta V_{\mathrm{Z}} = V_{\mathrm {fit}}$(RHCP)$-V_{\mathrm {fit}}$(LHCP), with the errors based on the accuracy of the fit Gaussian profiles, and the de-magnetized velocity that is assumed to be the mean of velocities of both polarized peaks: $V_{\mathrm{d}}=0.5\times(V_{\mathrm {fit}}$(RHCP)$+V_{\mathrm {fit}}$(LHCP)). Using the formula $\Delta V_{\mathrm{Z}}/B=$0.056 from \citet{baudry1997}, we calculated the magnetic field along the line of sight ($B_\mathrm{los}$), which is included in Figs.\,\ref{fig:g20p237}, \ref{fig:g24p148}, and \href{https://doi.org/10.5281/zenodo.14865259}{B.1}-\href{https://doi.org/10.5281/zenodo.14865259}{B.8}. The field is directed away from us when $V_{\mathrm{fit}}$(RHCP$)>V_{\mathrm{fit}}$(LHCP) and is directed toward us when $V_{\mathrm{fit}}$(RHCP$)<V_{\mathrm{fit}}$(LHCP). 

The bottom panels in Figs.\,\ref{fig:g20p237}, \ref{fig:g24p148}, and \href{https://doi.org/10.5281/zenodo.14865259}{B.1}-\href{https://doi.org/10.5281/zenodo.14865259}{B.8} show the circular and linear degrees of polarization of ex-OH emission. They were calculated as follows: $m_{\mathrm{c}}=V/I$ and $m_{\mathrm{l}}=P_{\mathrm{l}}/I$, respectively. The $V$ Stokes parameter, which is defined as 0.5$\times$(RHCP$-$LHCP), and the linearly polarized flux density, $P_{\mathrm{l}}$, which is defined as $(Q^2+U^2)^{0.5}$, are also presented. All spots with nonzero linearly polarized emission are summarized in Table \href{https://doi.org/10.5281/zenodo.14865259}{C.2} with their values of $I$, $Q$, $U$, $P_{\mathrm{l}}$, and $m_{\mathrm{l}}$, the electric vector PA (defined as $\chi_{\mathrm{l}}=0.5\times \arctan(U/Q)$, positive from north to east), $m_{\mathrm{c}}$, and the total polarization $m_{\mathrm{f}}=\sqrt{m_{\mathrm{l}}^2 + m_{\mathrm{c}}^2}$. 
The linear polarization vector is directed along the PA of the electric vector. The orientation of the magnetic field, $\Phi_{\mathrm{B}}$, is perpendicular to the linear polarization vector for $\sigma$ components. When the fraction of linear polarization exceeds 71\%, the amplification of $\pi$ component gain is high and so $\Phi_{\mathrm{B}}$ can become parallel (e.g., \citealt{fish2006_faraday}).

We calculated the brightness temperature ($T_{\mathrm B}$) as in \cite{wrobel99}, for the brightest maser in each source and transition. Since the maser spots are unresolved and we used the synthesized beam size as the area of the emission, $T_{\mathrm B}$ is a lower limit.
Below, we describe the individual sources in detail. 

\subsection*{G20.237$+$0.065 (hereafter G20)}
\begin{figure}[t!]
\centering
\includegraphics[scale=0.23]{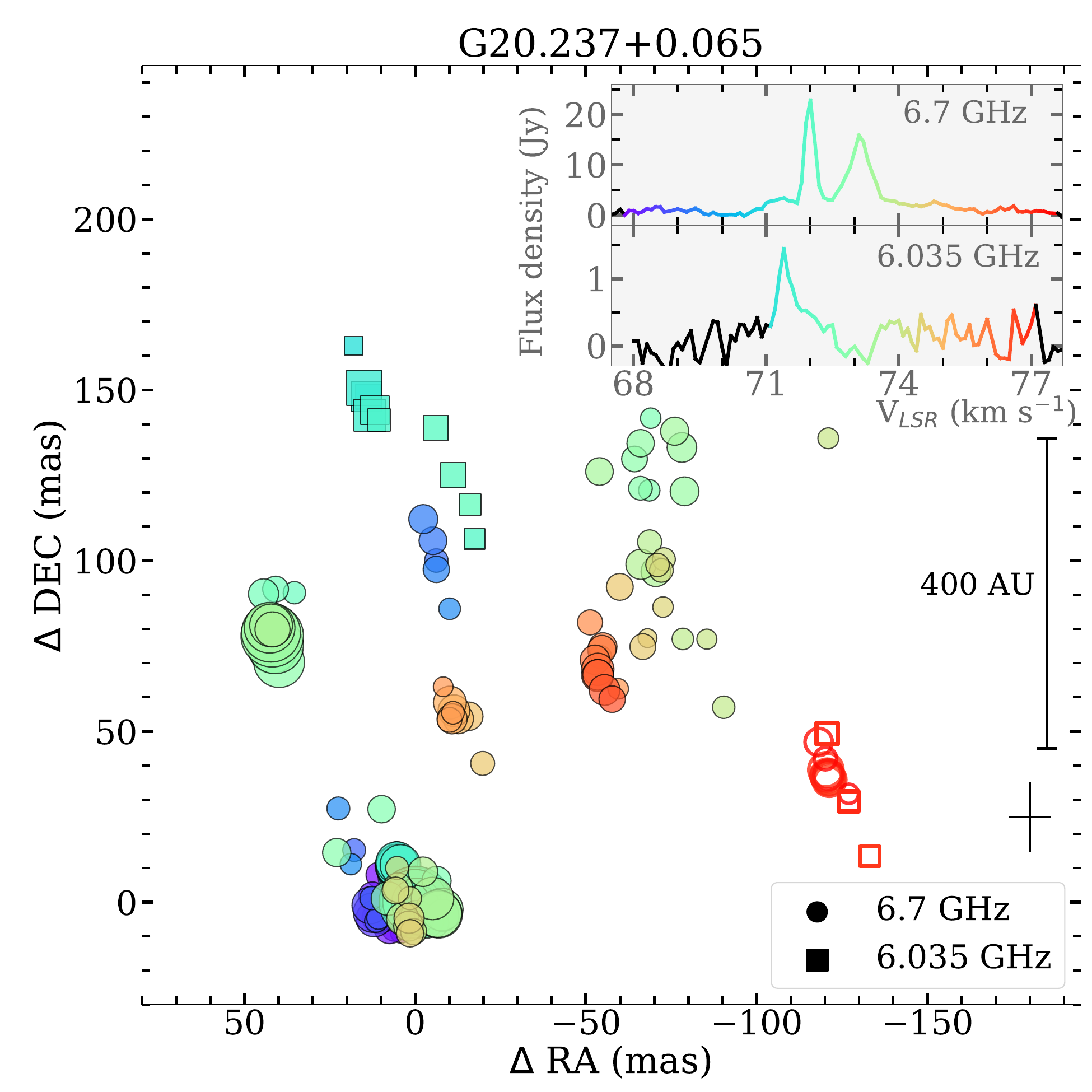}
\includegraphics[scale=0.23]{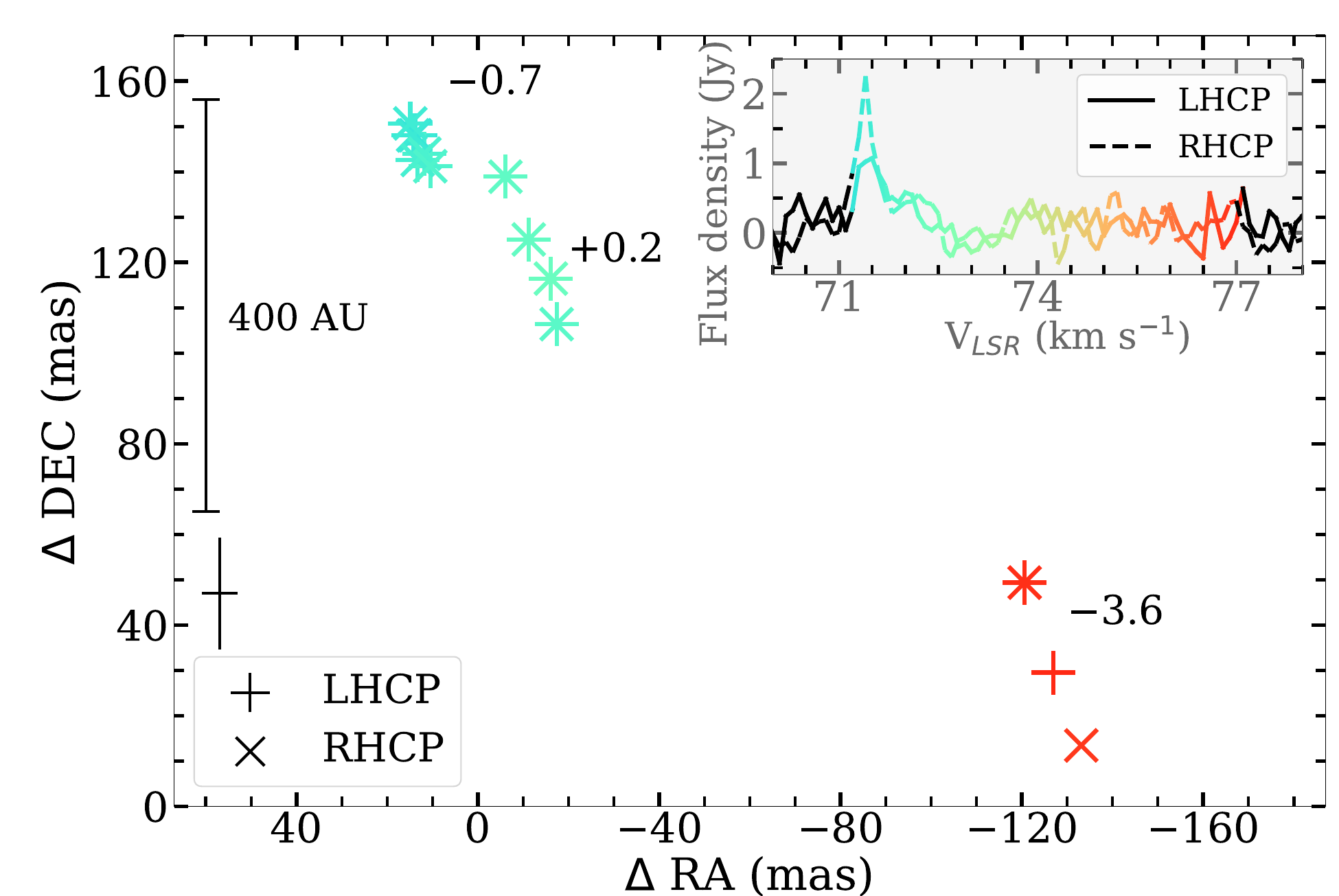}
\includegraphics[scale=0.23]{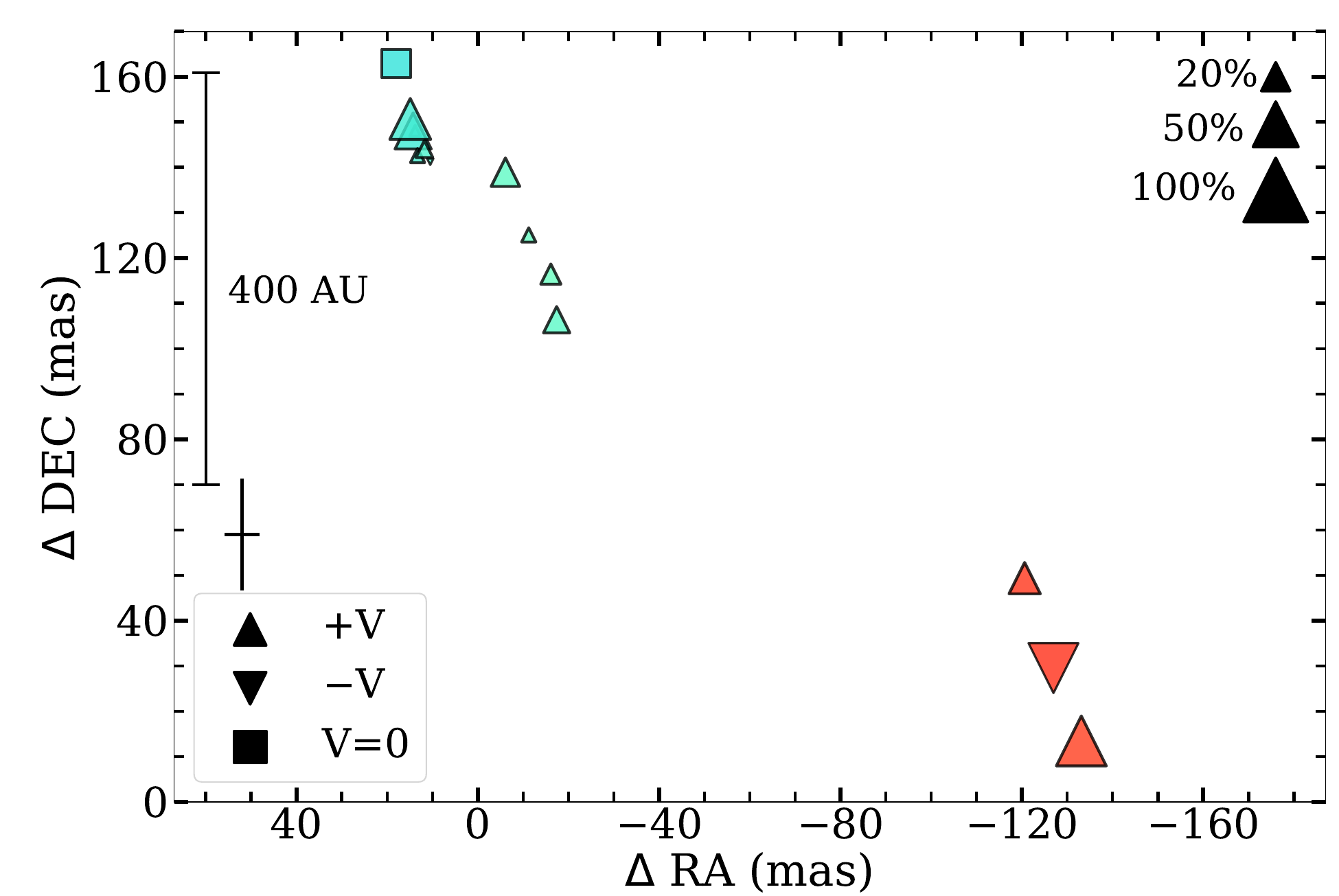}
\caption{Methanol and ex-OH maser emission in G20.237$+$0.065. {\bf Top:} Distribution of the 6.7 (circles) and 6.035~GHz (squares) Stokes $I$ maser spots. Symbol size is proportional to the square root of intensity, and the color corresponds to the LSR velocity, as is given in the total intensity spectra inserted at the top. LSR velocity ranges with no detections are shown in black.
%The parts of spectral profile marked in black color means there is no detection in that range of LSR velocities}. 
Non-filled symbols correspond to cases in which the coincidence of both transitions occurs. {\bf Middle:} Distributions of LHCP and RHCP of ex-OH maser spots. Numbers are the values of the line-of-sight magnetic field strength (in milligauss). {\bf Bottom:} Distribution of $V$ Stokes of ex-OH. Symbol size is proportional to the degree of circular polarization. The black cross indicates the maximum position error for the spot position.} %The black ellipse indicates the beam size.}
\label{fig:g20p237}
\end{figure}

%Rozmiar 6.7 (raxdec):  171.39786765690639  x  150.83899999999772  mas.
%Rozmiar 6.035 (raxdec):  151.15704415112708  x  149.44500000000005  mas.
The maser distributions of G20.237$+$0.065 are presented in Fig.~\ref{fig:g20p237}. The calculated near-kinematic distance is 4.41$\pm$0.39~kpc; the kinematic distance ambiguity was resolved by \cite{green2011} through H{\sc i} self-absorption.
%The central $V_{\rm LSR}$ of 72.6\kms corresponds to a  kinematic distance of 4.41$\pm$0.39\,kpc, with probability 100\%. 
% Cyganwoski et all 2013 : 4.38 kpc (+0.28,-0.3) - reid 2009
% Hu et all 2016 : 5.2$\pm$0.3 - from Green and McClure 2011 
The methanol maser emission is spread over 170\,mas$\times$150\,mas, corresponding to 750\,au$\times$660\,au. The morphology is complex, as was noted by \cite{bartkiewicz2016} based on EVN images. The ex-OH emission lies within a region of similar extent, 150\,mas$\times$150\,mas (660\,au$\times$660\,au), but this line is significantly weaker and has a simpler structure -- two groups of spots in the northeast and one in the southwest. The 6.7\,GHz transition covers a $V_{\rm LSR}$ range from 67.7 to 77.5\kms, and the 6.035\,GHz transition covers from 71.1 to 77.0\kms. We notice that in the most redshifted group, in the southwest, the two transitions coincide in velocity (within 0.5\kms) and position (within 34\,mas) and can be assumed to be coexisting in the same volume of gas, as was analyzed according to the procedure described in Sect.~2: 34~mas $\le$ 0.5(42~mas+33~mas).
%The ratio of maximum brightness temperature for methanol maser ($T_{\rm B}^6.7$) and ex-OH maser ($T_{\rm B}^6.035$), calculating according \cite{wrobel99}, is 6.95.  
The brightness temperature ($T_{\rm B}$) is at least 1.6$\times$10$^9$\,K for the methanol masers and 9.9$\times$10$^7$\,K for the ex-OH masers. 
%At that source, the methanol maser has a complex but irregular structure. A similar distribution was obtained during observations with smaller resolution using EVN by \citep{bartkiewicz2016}. A little different but also irregular distribution is shown in \citep{hu2016}. 

We identify three Zeeman pairs (Z$_1$-Z$_3$) in the ex-OH line, which indicate magnetic field strengths from $+$0.2 to $-$3.6\,mG (Table\,\href{https://doi.org/10.5281/zenodo.14865259}{C.1}). The magnetic field is directed toward the observer, with a possible reversal toward the middle group. 
%However, $\Delta V_{\mathrm{Z}}$ shows a significant uncertainty of 67\% (see Table~\ref{tab:zeemannpairs}. 
The degree of circular polarization is below 40\% for Z$_1$ and Z$_2$; for the redshifted pair (Z$_3$), we detect 29-74\% of circular polarization. We did not detect linear polarization from this source above a threshold of 34~mJy (3$\sigma$), %dla poli sigma jest mniejsza niż dla Q i U: 1sigma(poli)=11.36 mJy 
implying that the magnetic field lines may be oriented closer to the line of sight. 

There is another 6.7~GHz masing region 8\farcs6 toward the northeast, G20.239$+$0.065 (see Sect.~\ref{sec:couterparts}). We recovered the methanol emission similarly as \cite{bartkiewicz2016} but we have no detection of ex-OH maser above a threshold of 51~mJy~beam$^{-1}$ (3$\sigma$).

\subsection*{G24.148$-$0.009 (hereafter G24)}

\begin{figure}[h!]
\centering
\includegraphics[scale=0.21]{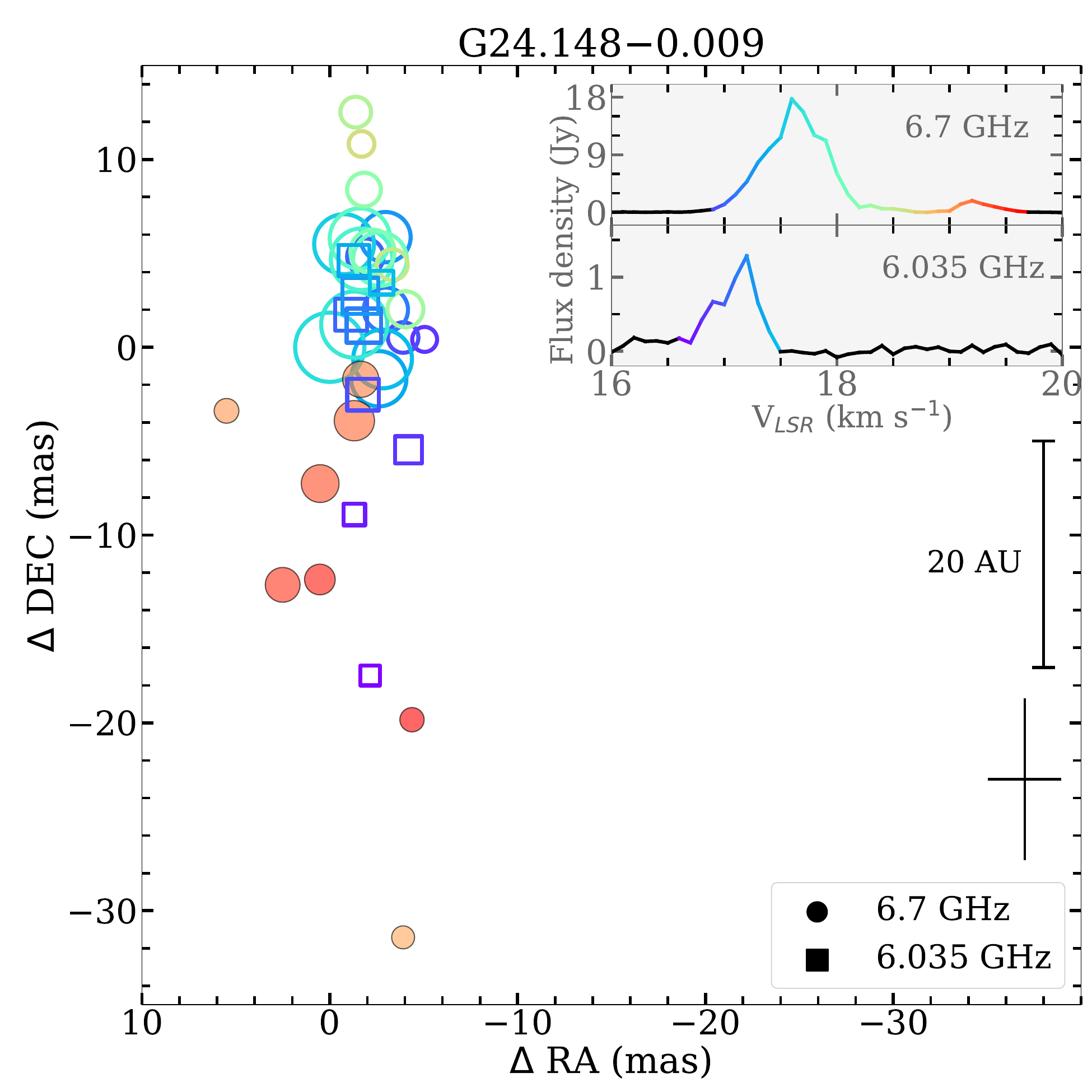}
\includegraphics[scale=0.21]{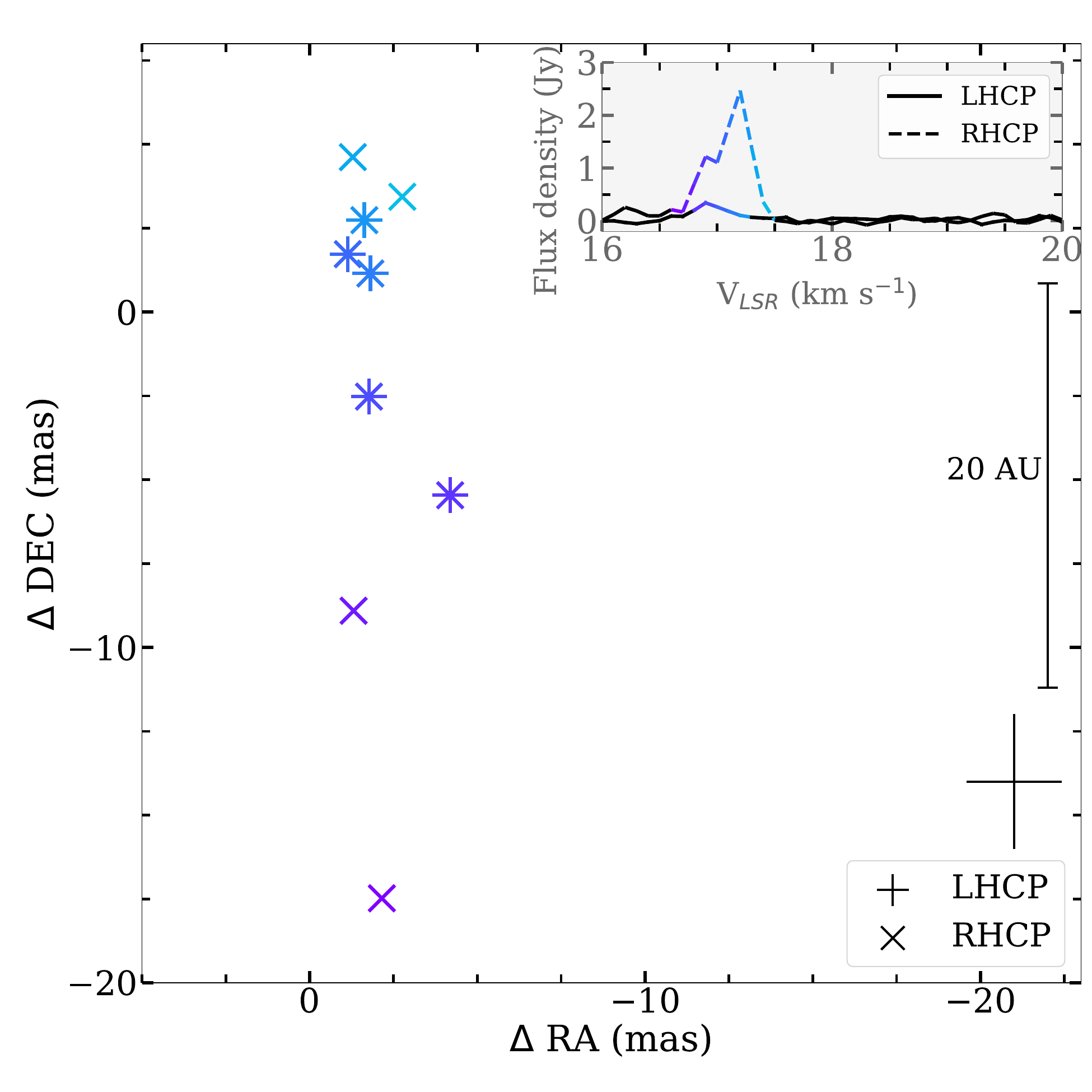}
\includegraphics[scale=0.21]{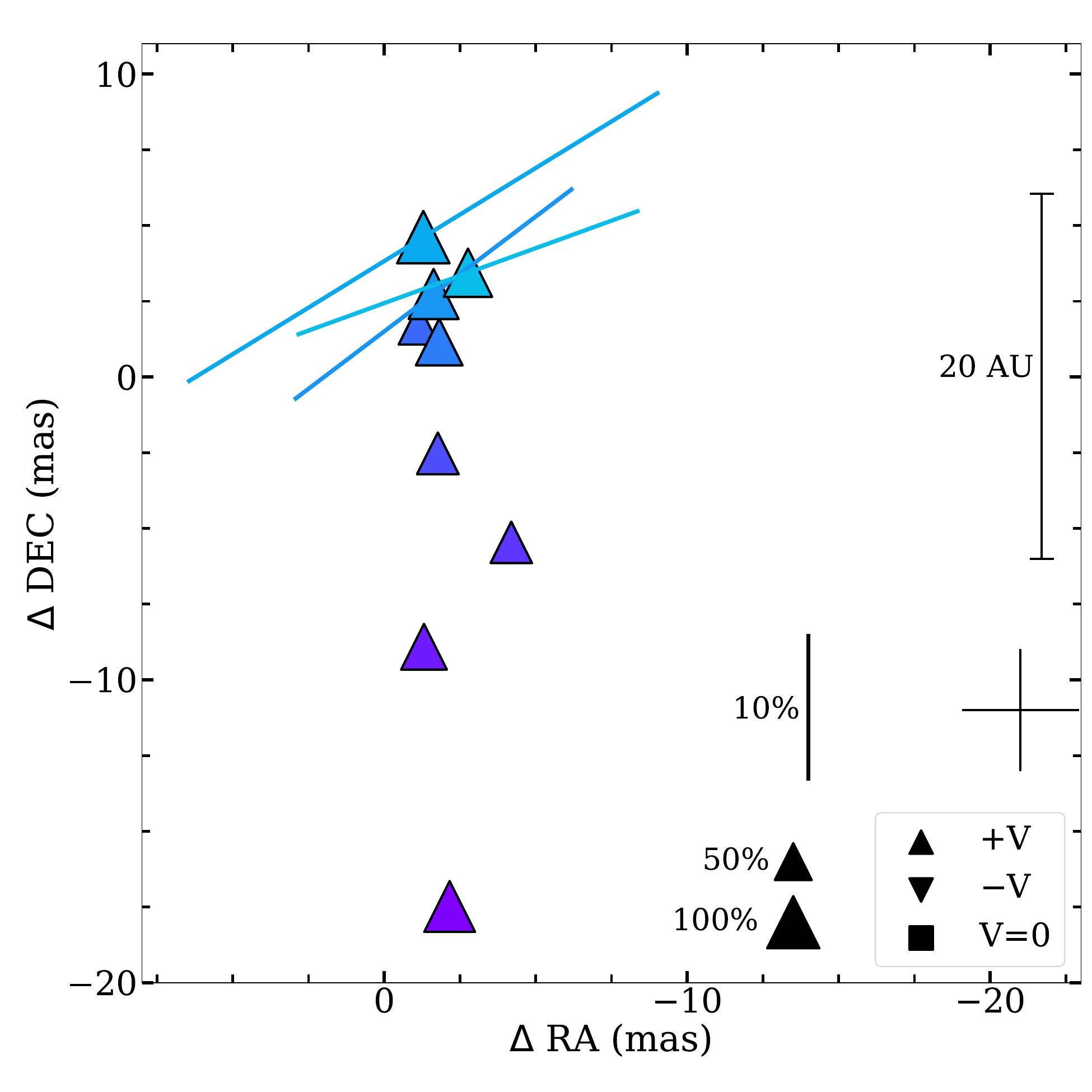}
\caption{Same as Fig. \ref{fig:g20p237} but for G24.148$-$0.009. In the bottom panel, the bars represent the direction of the planes containing the electric field vector. Their lengths are proportional to the percentage of linear polarization and their colors correspond to the LSR velocity, as is indicated in the inserted spectra in the top and middle panels.}
\label{fig:g24p148}
\end{figure}

The maser emission from both transitions is presented in Fig.\,\ref{fig:g24p148}. The near kinematic distance is 1.66$\pm$0.25\,kpc with a probability of 69\%, corresponding to the central velocity\footnote{We use the central velocity of masers, in cases where we do not have measurements of systemic velocities.} of 18.1\kms. The masing region is compact and elongated in the north-south direction. The centroids of the 6.7\,GHz spots occur within an area of 11\,mas$\times$44\,mas (18\,au$\times$73\,au) and those of the ex-OH spots within an area of 3\,mas$\times$22\,mas (5\,au$\times$37\,au). The emission covers the velocity ranges from 16.8\kms ~to 19.6\kms ~at 6.7\,GHz and from 16.6\kms ~to 17.4\kms ~at the 6.035\,GHz line. Coexistence of both lines occurs in the whole masing region except for the redshifted emission of the methanol line. 
%happens in a velocity of 1.2\kms and angular positions 22\,mas. 
The $T_{\rm B}$ is at least 1.1$\times$10$^9$\,K for methanol masers and 8.7$\times$10$^7$\,K for ex-OH masers. 

We are not able to identify Zeeman splitting due to a different number of spectral features in each polarization and attenuation or suppression of the LHCP feature, therefore estimation of the strength of the magnetic field is impossible. 
%In general spectra for LHCP and RHCP indicate the line of sight magnetic field is positive, directed away from us AK: po obliczeniu: B może być albo -1.1 mG albo +4 mG. 
The source is strongly polarized (bottom panel of Fig.~\ref{fig:g24p148}); $m_{\mathrm{c}}$ is from 59\% to 100\%, and three out of nine spots show $m_{\mathrm{l}}$ within a range from 24\% to 38\%. The directions of linear polarization vectors are from $-$52\degr~to $-$70\degr~(Table\,\href{https://doi.org/10.5281/zenodo.14865259}{C.2}), indicating a magnetic field oriented nearly northeast-southwest at $\Phi_{\mathrm{B}}$ = $+$30$\pm$6\degr.

\subsection*{G25.648$+$1.049 (hereafter G25)}
% ALMA - pobrane 
%{\color{blue} ALMA: project 2021.1.00311.S,  (0.27 asec res) but no image preview} 
The maser distribution is presented in  Fig.\,\href{https://doi.org/10.5281/zenodo.14865259}{B.1}. The near kinematic distance is 3.8$\pm$1\,kpc with a probability of 70\% for the central velocity of 41.4\kms. \cite{green2011} obtains a far kinematic distance of 12.5$\pm$0.4~kpc, derived from H{\sc i} self-absorption.
%from Green and McClure 2011: 12.5$\pm$0.4
%Bayandina 2019: 2.08 ± 0.37 - reid 2016
%Sunada 2007: 2.7 kpc

The regions of methanol and excited OH emission are clearly separated both in the spatial and in the velocity domain. Each transition forms three groups and ex-OH masers lie southward relative to the methanol masers. The 6.7\,GHz maser covers an area of 450\,mas$\times$325\,mas (1700\,au$\times$1225\,au and 5600$\times$4100~au for the near and far kinematic distances, respectively) and appears at a velocity range of 38.3-43.8\kms, whereas the 6.035\,GHz maser is spread over 300\,mas$\times$100\,mas (1130\,au$\times$377\,au and 3750$\times$1250~au for the near and far kinematic distances, respectively) and covers a velocity range of 38.3--39.9\kms. The smallest angular separation between these two transitions is $\sim$50\,mas. 
The $T_{\rm B}$ is at least 6.8$\times$10$^9$\,K for the methanol masers and 1.3$\times$10$^8$\,K for the ex-OH masers.   

Three Zeeman pairs are identified, showing a similar line-of-sight magnetic field strength from $-$6.2 to $-$7.0\,mG and indicating that the magnetic field is directed toward the observer (Table\,\href{https://doi.org/10.5281/zenodo.14865259}{C.1}). The emission shows strong circular polarization; 16 out of 31 spots are polarized above 80\%, and seven spots have $m_{\mathrm{c}}$ below 50\%. One spot has linear polarization with $m_{\mathrm{l}}$=95\% and a PA of $+$5\degr~(Table\,\href{https://doi.org/10.5281/zenodo.14865259}{C.2}), indicating a magnetic field directed east-west with $\Theta_{\mathrm{B}}= -85\pm2\degr$. We did not find strong evidence for the existence of a $\pi$ component (see Sect. \,\ref{sec:triplets}), so we assume that the linear polarization comes from $\sigma$ components and that $\Theta_{\mathrm{B}}$ is perpendicular to the linearly polarized vector.

\subsection*{G34.267$-$0.210 (hereafter G34)}
The results are presented in Fig.\,\href{https://doi.org/10.5281/zenodo.14865259}{B.2}. The near and far kinematic distances are 2.94$\pm$0.18\,kpc and 9.6$\pm$0.5\,kpc, with similar probabilities \citep{reid2019}.
%using Parallax-based Distance Calculator V2 did not get consistent outcome %6.12$\pm$3.09 with p=42%) 
%therefore we use value 10.6$\pm$0.3\,kpc from \citep{hu2016}.
The methanol maser emission is significantly more complex compared with ex-OH. It is spread over an area of
%Rozmiar 6.7 (raxdec): 123.33906808731466 x 218.5710000000043  mas.
%Rozmiar 6.035 (raxdec): 2.474549944836791 x 4.084999999996342  mas.
125\,mas$\times$220\,mas (370\,au$\times$650\,au and 1200\,au$\times$2100\,au for the near and far kinematic distances, respectively) %1325\,au$\times$2330\,au) 
and emission occurs in the LSR velocity range from 47.8 to 54.9\kms. The redshifted part is located in the east and coincides with the ex-OH maser spots %taking region 2.5\,mas$\times$4\,mas (8\,au$\times$12\,au and 24\,au$\times$39\,au for the near and far kinematic distances, respectively). 
within 40\,mas and 0.1\kms. The $T_{\rm B}$ is at least 4.1$\times$10$^8$\,K for the methanol masers and 1.5$\times$10$^8$\,K for the ex-OH masers.     

We are not able to identify any Zeeman pairs in the ex-OH transition. The LHCP emission is brighter but RHCP shows a more complicated spectrum. We notice a clear shift in velocity between these two polarizations, with RHCP being more positive; that indicates that the $B_\mathrm{los}$  is directed away from the observer and puts a lower limit on the magnetic field strength of 2.84~mG. Six out of eleven spots have $m_{\mathrm{c}}$ above 80\%, and seven spots show $m_{\mathrm{l}}$ from 6\% to 36\%. The planes of electric vectors rotate from one masing group to another over 50~au without any systematic changes. $\Phi_{\mathrm{B}}$ for the brightest four spots is $-$58$\pm$21\degr, while for three weaker it is $+$23$\pm$3\degr. Therefore, we cannot clearly state the direction of the magnetic field in this target.

\subsection*{G43.149$+$0.013 (hereafter G43)}
The ex-OH and methanol maser emission is presented in Fig.\,\href{https://doi.org/10.5281/zenodo.14865259}{B.3}. This target is part of the high-mass star formation region W49N lying at a distance of 11.11$^{+0.79}_{-0.69}$~kpc \citep{zhang2013}.   
%The far kinematic distance for this source is 10.96$\pm$0.35 kpc with 99\% probability, for central velocity 12\kms.  
%from Green and McClure 2011:11.4$\pm$0.5 kpc
%Rozmiar 6.7 (raxdec): 152.1320801904497 x 110.47200000000146  mas.
%Rozmiar 6.035 (raxdec): 258.7582846277395 x 249.16900000000197  mas.
In this source, the ex-OH masers are more complex than the methanol masers. The ex-OH emission is spread over 280\,mas$\times$250\,mas (3100\,au$\times$2800\,au) and over a LSR velocity range of 9.5-14.5\kms. It forms two groups elongated perpendicular to each other (the blueshifted emission is at a PA, from north to east, of $+$77\degr, while the redshifted emission is at $-$53\degr). The methanol masers appear at redshifted velocities from 13\kms ~to 14.2\kms, in three groups, similar to the appearance in VLBI images \citep{bartkiewicz2014}, over a region of 150\,mas$\times$110\,mas (1700\,au$\times$1200\,au). We notice the coincidence of both maser transitions in these three groups.
%Each of the three methanol emission groups coincides with exited OH transition, the red group within 6\,mas in angular position and to \kms in velocity, and the two orange groups within .    
The brightest, blueshifted part of the ex-OH transition appears not to have methanol maser counterparts. The $T_{\rm B}$ is at least 5.9$\times$10$^8$\,K for the methanol masers and 1.0$\times$10$^8$\,K for the ex-OH masers.  

We identify eight Zeeman pairs. The two most easterly ones imply values of the magnetic field strength above $+$3.0\,mG directed away from the observer, while in the remaining clumps the field is directed toward the observer with values from $-$1.3 to $-$6.5\,mG. That indicates the reversal of the magnetic field. Twenty out of fifty-four spots are circularly polarized above 80\% and twenty-two spots below 50\%. Eight spots (all belonging to three Zeeman pairs) show linear polarization, from 8\% to 48\%. The mean $\Phi_{\mathrm{B}}$ of the eastern part is $-$66$\pm$24\degr, while the electric vector of the redshifted linearly polarized ex-OH spot is 15\degr, indicating that $\Phi_{\mathrm{B}}$=$-$75$\pm$9\degr. Such differences in the spot distributions and the orientation of the magnetic field may indicate that these two regions are related to diverse kinematic regions.

\subsection*{G48.990$-$0.299 (hereafter G48)} 
The results are presented in Fig.\,\href{https://doi.org/10.5281/zenodo.14865259}{B.4}. The distance is 5.62$^{+0.59}_{-0.49}$~kpc, as was estimated by \cite{nagayama2015}.
%The kinematic distance is 5.44$\pm$0.47\,kpc with probability equal 100\% for central velocity 69.1\kms.
%from Green and McClure 2011: brak
Again, in this target, the ex-OH emission is more complex than the methanol masers. 
The methanol maser emission shows a simple structure, two spots at  LSR velocities of 71.5\kms ~and 71.6\kms. The ex-OH masers appear in a wider velocity range from 66.7 to 69.6\kms ~and cover a region of 290\,mas$\times$135\,mas, which corresponds to 1630\,au$\times$760\,au. They are elongated along a line from southeast to northwest. The $T_{\rm B}$ is at least 1.9$\times$10$^8$\,K for the methanol masers and 2.0$\times$10$^8$\,K for the ex-OH masers. \cite{nagayama2015} report water maser emission at 22~GHz at LSR velocities from 65\kms ~to 68\kms ~showing north-south elongation and proper motions implying expansion. 
%{\bf can we check the coordinates and see where meth/water lie? {\color{red}AK: I added position of water maser to the Fig. \ref{fig:g48_ir}.}
%https://watermark.silverchair.com/pasj_67_4_65.pdf }

The ex-OH transition forms four Zeeman pairs. Two western groups have positive values of $B_\mathrm{los}$ ($+$2.7 and $+$2.3\,mG), directed away from the observer, and two eastern groups, separated by $\sim$250\,mas, have negative values ($-$0.7 and $-$4.8\,mG) of $B_\mathrm{los}$, indicating a magnetic field directed toward the observer. Circular polarization is below 50\% for 18 out of 29 spots and above 80\% for 5 out of 29 spots. Linear polarization is detected for six spots in the two brightest groups, with degrees 8-31\%. The electric vectors are consistent over the whole region with PAs from $-$17 to $-$56\degr, indicating $\Phi_{\mathrm{B}}$= +$54\pm$7\degr. Again, we see a clear reversal of the line-of-sight magnetic field. 

\subsection*{G49.490$-$0.388 (hereafter G49)}
This target belongs to the well-known high-mass star formation region W51. 
The ex-OH and methanol maser emission is presented in Fig.\,\href{https://doi.org/10.5281/zenodo.14865259}{B.5}. The distance is 5.41$^{+0.31}_{-0.28}$~kpc, based on the trigonometric parallax measurement using water masers by \cite{sato2010}; we note that the result of 5.1$^{+2.9}_{-1.4}$~kpc using 12.2~GHz methanol masers, by \cite{xu2009par}, is less accurate. 
%5.48$\pm$0.30\,kpc with probability 90\% and for central velocity 55.0\kms.
%from Green and McClure 2011: 5.1 $\pm$ 2.9 kpc
%Rozmiar 6.7 (raxdec):  465.50577326036836  x  596.9600000000028  mas.
%Rozmiar 6.035 (raxdec):  267.9935202726133  x  531.0640000000005  mas.
The 6.7~GHz methanol maser emission covers an area of 0\farcs47$\times$0\farcs55, corresponding to 2550\,au$\times$3000\,au, and appears in two velocity ranges, from 51.2 to 52.3\kms ~and from 57.5 to 60.2\kms. The emission is very bright, with a maximum flux density of 684~Jy. The morphology agrees well with the distribution derived from EVN observations, but it is more extended. \cite{fujisawa2014} reported a smaller masing region of 0\farcs37$\times$0\farcs31.
However, \cite{etoka2012} imaged emission over 3\arcsec$\times$2\arcsec. Ex-OH masers are weaker with a maximum flux density of 7~Jy and are located in the west over a region of 0\farcs27$\times$0\farcs53 (1460\,au$\times$2870\,au)  at intermediate LSR velocities from 51.6 to 58.3\kms. Like \cite{etoka2012}, we do not find any spatial overlap between the two transitions. The $T_{\rm B}$ is at least 5.9$\times$10$^8$\,K for the methanol masers and 1.0$\times$10$^8$~K for the ex-OH masers. 

We identified nine Zeeman pairs. Z$_1$-Z$_3$ are new ones compared with the results of \cite{etoka2012} (their Table~3). For all groups, the field is directed away from the observer, and its values are from $+$3.3\,mG to $+$8.1\,mG, which is consistent with \cite{etoka2012}. The emission is strongly polarized; 41 of 85 spots have a degree of circular polarization above 80\% and only 20 spots below 50\%. Twenty spots in four Zeeman pairs show linear polarization in the range 5-22\%. The mean PA of the electric vectors of 0$\pm$10\degr indicates the magnetic field oriented at $-$90\degr. However, we notice different values of $\chi_{\mathrm{l}}$ for each maser group, similar to in G34.

\subsection*{G69.540$-$0.976 (hereafter G69)}
%{\color{blue} ALMA: project 2021.1.00311.S  bright continuum emission detected north east of maser, but some continuum emission seen also on methanol position. also some lines seen.}
This target is associated with the high-mass star formation region Onsala\,1 (ON1). The maser distributions are presented in Fig.\,\href{https://doi.org/10.5281/zenodo.14865259}{B.6}. The distance based on the trigonometric parallax is 2.57$^{+0.34}_{-0.27}$~\,kpc \citep{rygl2010}.
%4.01\,kpc with p=0.69, vcentr=13.4\kms - bessel
%from Green and McClure 2011:brak
%Rozmiar 6.7 (raxdec): 473.6226838742033 x 1089.8630000000012 mas.
%Rozmiar 6.035 (raxdec): 1047.6477861530684 x 1079.467000000001 mas.
%6.7\,GHz and 6.035\,GHz masers is observed: 
%(1) for the blue-shifted features away 50\,mas and within 0.9\kms in the velocity domain, 
%(2) for the two of southern, red-shifted groups within 90\,mas and 0.9\kms in the velocity domain. 
The methanol masers cover an area of 0\farcs47$\times$1\farcs09, or 1220\,au$\times$2800\,au, over two ranges of  LSR velocities, from $-$0.5\kms ~to 2.6\kms ~and from 14.1\kms ~to 15.7\kms. The observed spatial distribution of the 6.7 GHz masers is more complex compared to images obtained using EVN in 2006-2008 by \cite{rygl2010}, since these authors reported only two groups with an angular separation of 0\farcs94. %and by 15\kms. 
However, our results agree well with the EVN images taken in 2015 by \cite{surcis2022}, in which three groups were reported similarly as by \cite{sugiyama2011} using JVN in 2006-2008. The ex-OH masers are spread over 1\farcs05$\times$1\farcs08, i.e. 2690\,au$\times$2775\,au, and also come from two velocity ranges: from $-$1.6\kms ~to 2.3\kms ~and from 12.2\kms ~to 15.2\kms. 
We identify two regions where the overlap of both transitions is seen: in the north, the blueshifted masers coincide within 46~mas (33.8\,mas in RA and 31.6\,mas in Dec); and in the south, the redshifted masers coincide within 28~mas (23.6\,mas in RA and 13.6\,mas in Dec). 
The $T_{\rm B}$ is at least 4.4$\times$10$^9$\,K for the methanol masers and 3.4$\times$10$^8$\,K for the ex-OH masers. 

We identify five Zeeman pairs in the ex-OH transition, at redshifted velocities, implying a negative magnetic field (directed toward us): $B_\mathrm{los}$ is from $-$1.2\,mG to $-$6.2\,mG. The same five pairs were reported by \cite{green2007}. We are not able to identify the Zeeman splitting for the blueshifted emission due to different LCHP and RCHP spectral features. The degree of circular polarization is below 50\% for 59 of 96 spots and above 80\% for 38 spots. In each Zeeman pair, there are spots (in total 26) that are linearly polarized with degrees from 8\% to 48\%. The means of $\Phi_{\mathrm{B}}$ are: $-50\pm17\degr$ for the northern and blueshifted region and $+80\pm12\degr$ for the southern and redshifted region. \cite{green2007} reported two ex-OH features with $m_{\mathrm{l}}$ of 18.5\% and 10\% and $\chi_{\mathrm{l}}$ of $-87.7\degr\pm2\degr$ and $-42.5\degr\pm0.7\degr$, respectively. They correspond to the
%Zeeman pairs Z$_4$ and Z$_3$ to 
spots at the LSR velocity of 14.3--14.5\kms~ listed in Table\,\href{https://doi.org/10.5281/zenodo.14865259}{C.2}. We note that \citet{surcis2022} estimated a mean magnetic field direction of $-51\pm69\degr$ from linearly polarized methanol masers.
%Again, the electric vectors are not well-ordered in this target similarly as in G34 and G49.

\subsection*{G81.871$+$0.781 (hereafter G81)}
%{\color{blue} ALMA: No continuum emission in project 2021.1.00311.S, but we should check Gomez et al. 2023 for SiO emission seen by ALMA.}
This target is in the well-known high-mass star formation region W75N. Its distance is 1.30$\pm$0.07\,kpc \citep{rygl2012}. The methanol and ex-OH maser distributions are presented in Fig.\,\href{https://doi.org/10.5281/zenodo.14865259}{B.7}. 
%2.18 kpc with p=61\%, vcent=6.1\kms.
%from Green and McClure 2011:brak
The methanol masers are spread over an area of 0\farcs49$\times$1\farcs065, or 638\,au$\times$1384\,au, and appear in the $V\mathrm{_{LSR}}$ range of 2.6--9.7\kms. That is consistent with EVN results from \cite{rygl2012}. Ex-OH maser emission is less complex; it is located in the central part of the 6.7\,GHz emission over an area of 59\,mas$\times$312\,mas (77\,au$\times$406\,au), over a $V\mathrm{_{LSR}}$ range of 6.2--9.2\kms. 
%Rozmiar 6.7 (raxdec): 490.53250466656067 x 1064.3719999999987  mas.
%Rozmiar 6.035 (raxdec): 59.0260432264342 x 312.08999999999776  mas.
It appears as three groups. We notice an overlap between
the two masers in the southern group: two spots coincide within 21\,mas (15.3\,mas in RA and 14.7\,mas in Dec). Similarly, in the northern group, six spots coincide within 65\,mas (22\,mas in RA and 61.6\,mas in Dec), and in the centrally located group, nine spots coincide within 61~mas (33.7\,mas in RA and 50.5\,mas in Dec). 
The $T_{\rm B}$ is at least 14.3$\times$10$^9$\,K for the methanol masers and 2.1$\times$10$^8$\,K for the ex-OH masers. 

We report three Zeeman pairs,
%where calculating the value of the line of sight magnetic field was impossible because of the different amounts of Gauss profiles in the spectra of LHCP and RHCP. 
for which the $B_\mathrm{los}$ is positive and directed away from the observer with values of $+$2.3\,mG for the southern group and of $+$7.3 and $+$8.5\,mG for the central pairs. In the northern group, we are not able to identify LHCP and RHCP features belonging to a likely Zeeman pair. The degree of circular polarization is above 80\% for 16 of 39 spots and below 50\% for 14 spots. Fourteen spots show linear polarization, the majority of which have a degree from 4\% to 15\%. Three redshifted spots at $V_{\rm{LSR}}$ 8.3--8.6\kms ~are much more polarized; their degrees are 38--58\%. The mean value of the PAs of the electric vector planes is 29$\pm$13\degr, indicating a magnetic field directed approximately along $-61\degr$. Again, the source shows diverse directions of the magnetic field, possibly reflecting changes in the gas kinematics. 

\subsection*{G108.766$-$0.986 (hereafter G108)} 
The distributions of both methanol and ex-OH masers are compact (8\,mas$\times$11\,mas and 7\,mas$\times$30\,mas, respectively) and well separated {in position (by around~80~mas). These are presented in Fig.\,\href{https://doi.org/10.5281/zenodo.14865259}{B.8}. 
%\cite{hu2016} confirm such distribution for methanol masers. 
The near kinematic distance is 2.81$\pm$0.23\,kpc, with a probability of 72\% for the central velocity of $-$45.6\kms, implying a spatial separation between the two maser transitions of 225~au. 
%from Green and McClure 2011:break
Both lines appear at a similar LSR velocity range, from $-$46.4\kms ~to $-$44.8\kms ~for the methanol maser and from $-$46.1\kms ~to $-$44.8\kms ~for the ex-OH maser line. The smallest angular separation between the two masers is 67\,mas, which exceeds twice the $\gamma$ parameter, and we assume no overlap of both transitions. 
The $T_{\rm B}$ is at least 8.3$\times$10$^8$\,K for the methanol masers and 6.3$\times$10$^7$\,K for the ex-OH masers.
%Rozmiar 6.7 (raxdec):  7.989227935576119  x  10.54200000000094  mas.
%Rozmiar 6.035 (raxdec):  7.281189209269179  x  29.363000000000028  mas.

We identify two Zeeman pairs and we estimate the value of the magnetic field as $-$8.5\,mG and $-$10.5\,mG, indicating that the $B_\mathrm{los}$ is directed toward the observer. The ex-OH emission consists of 12 spots, of which three are more than 80\% circularly polarized and two are less than 50\% polarized. Only one spot shows linear polarization with a degree of 12\%, indicating the value of $\Theta_{\mathrm{B}}=+22\pm9\degr$.

\section{Discussion} \label{sec:discussion}
\subsection{Counterparts} \label{sec:couterparts}
First, we discuss the counterparts of HMYSOs at other frequencies to derive information on the HMYSO environment on a larger scale. Centimeter radio emission identifies thermal jet or UCH{\sc{ii}} regions and, thus, a more evolved stage of the HMYSO.  
The earliest phases are seen in the ALMA dust continuum emission at 1 and 3\,mm, which identifies star-forming cores, disks, and envelopes. To trace outflows, and in particular the outflow direction, which is interesting to compare to the magnetic field orientation, we used 22\,GHz water masers and archival ALMA\footnote{The ALMA Science Archive is available at \url{https://almascience.org/aq/?result_view=observations}} spectral line data of SiO, SO, and the CO isotopologs in ALMA Band 6. 
The SiO, SO, and water masers are all shock tracers, while the CO isotopologs ($^{12}$CO, $^{13}$CO) trace large-scale gas motions. We typically used $^{13}$CO because it is more widespread and brighter than the (often) optically thin C$^{18}$O transition but not as optically thick (and complex) as the $^{12}$CO line. For warm gas, CH$_3$CN and HC$_3$N can be used as disk tracers, and 
the CH$_3$CN $K$ ladder to obtain rotational temperatures and H$_2$CO along with thermal methanol to trace young star-forming object (YSO)-heated gas.
To constrain the clump properties, we used {\em Spitzer} 3.5, 4.5, 6, 8, and 24\,$\mu$m, and {\em Herschel} 70\,$\mu$m data that trace the warm dust heated by the central object(s). In Table\,\href{https://doi.org/10.5281/zenodo.14865259}{C.3}, we list the HMYSOs' positions from the Hi-GAL catalog \citep{molinari2016, molinari2016cat, elia2021}. The Herschel astrometric accuracy is $\sim$2~arcsec \citep{molinari2016}. The target positional accuracy of \textit{Spitzer} is about 1~arcsec, and the angular resolution is $\sim$2~arcsec\footnote{\label{caltech}\url{https://irsa.ipac.caltech.edu/}}. The detailed descriptions of counterparts of all sources along with figures are in Appendix \href{https://doi.org/10.5281/zenodo.14865259}{D}.

\subsection{Evolutionary stage}
\label{sec:evolve}
%{\color{red} KR: I would first write that 70um emission is indicate of warm dust, heated by the protostar. As the protostar evolves, the 70um emission increases in strength.}  
As a tracer of warm dust, the far IR 70\,$\mu$m emission increases in strength as the protostar evolves.  
Table\,\href{https://doi.org/10.5281/zenodo.14865259}{C.3} shows the strongest 70\,$\mu m$ emission for G49 and for G43, G25, and G48. That suggests that those objects are the most evolved in our sample. G20, G24, and G34 show weak 70$\,\mu$m emission, indicating an earlier evolutionary phase, but the ratio $S_{70~\mu\mathrm{m}}/S_{24\,\mu\mathrm{m}}$ suggests that only G34 is an early HMYSO. We do not have information about 70\,$\mu$m emission for G69, G81, and G108.
Another evolutionary indicator is $\frac{L_{\mathrm{bol}}}{M}$ \citep{molinari2008}. Higher values indicate a later evolutionary phase. This criterion also supported G43 as the most evolved source, followed by G25; these sources are also characterized by the highest average dust temperatures in our sample, of 39 and 40\,K, respectively. 
Based on the $\frac{L_{\mathrm{bol}}}{M}$ indicator, G20 and G34 are the youngest sources, and G24 and G48 are slightly more evolved. All those sources have T$_D$ in the range between 21 and 25~K. 
We do not have information about $\frac{L_{bol}}{M}$ and T$_D$ for G49, G69, G81, and G108. 
Both indicators, IR 70\,$\mu$m emission as well as $\frac{L_{\mathrm{bol}}}{M}$, concern the evolutionary states of clumps, but these can host more than one core or YSO. A better evolutionary indicator for a single YSO is the existence of thermal jets or a UCH{\sc ii} region. The sources with the strongest UCH{\sc ii} emission are G69, G43, and G48. In the case of G20 and G24, we did not find any confirmation as to whether observable radio continuum emission is related to a UCH{\sc ii} region or to a jet. However, the $S_{70~\mu\mathrm{m}}/S_{24\,\mu\mathrm{m}}$ ratio indicates that both sources are in late evolutionary phases, so we can assume that the radio continuum emission is related to UCH{\sc ii} emission, rather than to a thermal jet. G81 is associated with the radio continuum VLA1, which was identified as a radio jet. Also, G25 shows evidence that the radio continuum is associated with the radio jet. G34 and G108 do not have observable radio continuum emission. G49 is close to the UCH{\sc ii} but the maser emission can be related to the eastern core, which is the dominant accretion source in that region \citep{shi2010outflow} and where the radio continuum was not detected \citep{gaume1993}.  
We further analyze the source evolutionary phase in Sect.~\ref{sec:coin}.

\subsection{Polarization properties and magnetic field}
\label{sec:polarization}
\subsubsection{Zeeman pairs}
\label{sec:polarizationZP}
In total, we have identified 37 Zeeman pairs in eight HMYSOs: G20, G25, G43, G48, G49, G69, G81, and G108. We list the observed and fit peak flux densities and other parameters for all Zeeman pairs in Table\,\href{https://doi.org/10.5281/zenodo.14865259}{C.1}.
Toward G24, we are not able to identify any Zeeman splitting due to the weakness of LHCP emission, and similarly, toward G34, we do not report any Zeeman splitting since it is impossible to define clearly the RHCP feature related to the single LHCP feature appearing in the ex-OH emission. The calculated magnetic field along the line of sight is from 0.2~mG up to 10.6~mG, with mean and median values of 4.8~mG and 5.2~mG, respectively. These values are typical for high-mass star-forming regions (e.g., \citealt{bartkiewicz2005}, \citealt{fish2005}, \citealt{vlemmings2008}, \citealt{green2015}). However, \cite{surcis2022} report slightly higher |$B_\mathrm{los}$|, in a range from 9~mG to 40~mG in their sample of 31 targets, based on methanol masers' 6.7~GHz transition. 

The direction of $B_\mathrm{los}$ is consistent for all measurements for five sources. The magnetic field is directed toward the observer in G25, G69, and G108 and away from the observer in G49 and G81. In the case of G43 and G48, we observe magnetic field reversals similar to the one reported for Cep\,A \citep{bartkiewicz2005}. This may indicate a toroidal component of the magnetic field related to a disk. We do not notice any dependence between the size of the masing region and the magnetic field reversals. G43 and G48 are sampled by the Zeeman pairs over regions of 2000-3600~au, similar to G69, which shows the magnetic field directed to the observer over the whole region.

The relationship between the strength of the magnetic field and gas number density in molecular clouds in which the contraction is driven by ambipolar diffusion is defined as $|B|\sim n^{0.47}$ \citep{crutcher1999}. We  
%estimated the gas density 5.6$\times$10$^5$\,cm$^{-3}$ for B$=0.5$\,mG in G33.103$+$0.108 and 15.9$\times$10$^5$\,cm$^{-3}$ for B$=0.8$\,mG in G349.09$+$0.11 based on {\bf ???}. Using this data we 
estimated the gas density following \cite{bayandina2014}, considering $B_\mathrm{los}$ in the range of 0.2-10.6~mG, which leads to derived gas densities ($n_\mathrm{H_2}$) in the range from 0.1 to 378$\times$10$^6$~cm$^{-3}$.
%The results can be understated depending on the strength linear polarization and the resulting component magnetic field perpendicular to the line of sight.
%(Blos/0.5)^(1/0.47)*5.6*10^5
%(Blos/0.8)^(1/0.47)*15.9*10^5
%rho=srednia z obu równań
\subsubsection{Sky-plane B orientation}
In nine HMYSOs, we found linearly polarized ex-OH masers: G24, G25, G34, G43, G48, G49, G69, G81, and G108. The linear polarization percentage is generally low, less than 40\% for the majority of spots, and the mean and medium are 20\% and 15\%, respectively (Fig.~\ref{fig:hist_ml}). G49 and G69 show the most numerous linearly polarized spots (more than 20).%In two sources, G25 and G108, we found only one maser spot with $P_{\mathrm{l}}$ of 53~mJy~beam$^{-1}$ and 108~mJy~beam$^{-1}$, respectively. 
In two sources, G25 and G108, we found only one maser spot with $m_{\mathrm{l}}$ of 95\% and 12\%, respectively. In G25, a linearly polarized ex-OH maser spot indicates that the plane of the magnetic field is $-85\pm2$\degr if the emission is associated with a $\sigma$ component or $+5\pm2$\degr if the emission is associated with a $\pi$ component. In both cases, the magnetic field is not parallel or perpendicular to the outflow.

\begin{figure}
    \centering
    \includegraphics[width=\columnwidth]{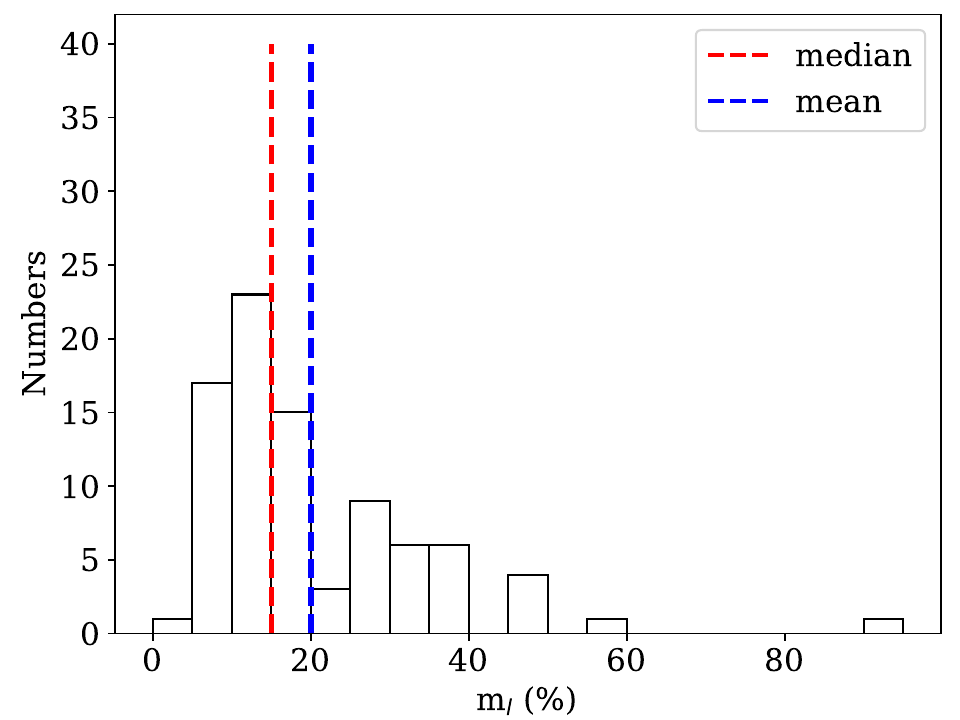}
    \caption{Histogram of the percentage of linear polarization of all spots with detected linearly polarized emission. They are listed in Table\,\href{https://doi.org/10.5281/zenodo.14865259}{C.2}.}
    \label{fig:hist_ml}
\end{figure}

In G20, we did not find any linearly polarized emission above a threshold of 2$\sigma$, implying that the magnetic field lines may be oriented closer to the line of sight.
In G24 and G48, the electric vectors are well ordered, as are the vectors for the most blueshifted emission from G43.
That indicates a consistent large-scale magnetic field orientation in the sky plane. Similar well-ordered linear polarized vectors were noticed by \cite{surcis2013, surcis2015, surcis2019, surcis2022} from observations of bright 6.7~GHz methanol masers.

In G34, G49, G69, and G81, and the more redshifted parts of G43, the directions of electric vectors are strongly variable, and changes do not correlate with spot velocities. 
\cite{koch2018} used ALMA dust polarization observations to estimate the direction of the magnetic field at different angular scales in G49 (W51e2e): at a resolution of 2~arcsec, the magnetic field is directed almost east to west, while at the higher angular resolution it is radially distributed, suggesting that the field is being dragged by the mass inflow. Ex-OH masers, lying offset from the young star and indicating a mean magnetic field oriented from east to west, fit well into this picture by tracking the large-scale magnetic field. \cite{etoka2012} inferred that the ex-OH was part of a flow in which the dynamics were magnetically dominated.

The magnetic field in G81, estimated from 18\,cm OH masers by \cite{hutawarakorn2002}, reverses its direction on opposite sides of the disk, indicating a toroidal component of the magnetic field in the disk. \cite{surcis2011} analysed the 22\,GHz water masers associated with the VLA\,1 and VLA\,2 radio continuum sources. They reported a tightly ordered magnetic field around VLA\,1 (aligned with the large-scale molecular outflow) and revealed an ordered magnetic field around VLA\,2 (not parallel to the outflow). These strong magnetic fields, 0.7~G and 1.7~G, around VLA\,1 and VLA\,2, respectively, are related to shock compression of the gas. It is not easy to compare our results with the previous observations due to the time difference and proper motions of masers, which might be significant at this relatively close distance. We do not detect such a magnetic field reversal, since the ex-OH masers trace a more compact region than the 18~cm OH masers (2~arcsec). The magnetic field strength and orientation (away from the observer) derived from Zeeman splitting of ex-OH is consistent with measurements of the Zeeman splitting of 18~cm OH (see Fig.~11 of \citealt{hutawarakorn2002}). 

Table~\ref{tab:papa} lists the mean directions of the magnetic fields estimated using linearly polarized ex-OH masers ($\Phi_{\mathrm{B}}$), methanol and ex-OH elongation directions (PA$_{\mathrm{6.7}}$ and PA$_{\mathrm{6.035}}$), estimated from least-square fitting to the spot distributions. These are compared with the directions of outflows (PA$_{\mathrm{out}}$) where available (see Sect.~4.1). Based on the Kolmogorov-Smirnov test, we cannot find any correlation for the linear distributions of both masers with the outflows and with the estimated magnetic field orientations in the sky plane as well as between outflows and magnetic field vectors themselves. Even for the targets with the most elongated maser structures, such as G24, G43, G48, or G81, we still cannot sketch a clear scenario relating to maser structures and the magnetic field. This is similar to the conclusions of \cite{surcis2013}. 
However, in \cite{surcis2022}, the summary for a whole sample of 31 targets reports a bimodal distribution, with: 1) half the magnetic field directions being perpendicular, and 2) the other half being parallel to the outflow. It is difficult to study dependencies between sub-arcsecond maser data and images with a larger angular scale searching for outflows and disks.

\begin{table*}
\centering
\caption{Comparisons of positions angles of the magnetic field, methanol maser distributions, ex-OH distributions, and the PAs of outflows.}
\begin{tabular}{ccccccccc}
\hline
 Source & $\Phi_{\mathrm{B}}$ & PA$_{\mathrm{6.7}}$ & PA$_{\mathrm{6.035}}$ & PA$_{\mathrm{out}}$ & |PA$_{\mathrm{6.7}}$-PA$_{\mathrm{6.035}}$| & |$\Phi_{\mathrm{B}}$-PA$_{\mathrm{out}}$| & |$\Phi_{\mathrm{B}}$-PA$_{\mathrm{6.7}}$|& |$\Phi_{\mathrm{B}}$-PA$_{\mathrm{6.035}}$| \\
   & (\degr) & (\degr) & (\degr) & (\degr) & (\degr) & (\degr) & (\degr) & (\degr) \\
\hline
 G20.237 & - & $-$70 & $+$50 & - & 60 & - & -  & -  \\
 G24.148 & $+$30$\pm$6 & $-$1 & $+$2 & $+$90 & 3 & 60 & 21 & 28 \\
 G25.648 & \textit{$-$85$\pm$2} & $-$72 & $+$84 & $-$30 & 24 & \textit{55} & \textit{13} & \textit{11} \\
 G34.267 & $+$80$\pm$23 & $+$8 & $+$24 & - & 16 & - & 72 & 56 \\
% G43.149 & $-$67 & $-$54 & $-$60 & & 6 & & 13\\
G43.149  & $-$66$\pm$24/$-$75$\pm$9$^*$ & -/$-$54 & $+$77/$-$53$^*$ & \textit{$+$40} & -/1  & \textit{106/115}& -/21 & 37/22\\ % 1st region blue and green spots, 2nd region red and orange spots
 G48.990 & $+$54$\pm$7 & {\it $+$44} & $-$69 & $+$20$^{(1)}$ & 67 & 30 & {\it 10} & 57 \\
 G49.490 & $-$90$\pm$10 & $+$41 & $+$10 & $-$46$^{(2)}$ & 31 & 44 & 49 & 80 \\
% G69.540 &  $-$83 & $+$29 & $+$89\\
 G69.540 &  $-$50$\pm$17/$+$80$\pm$12$^{**}$ & $+$39/$-$68$^{**}$ & $+$47/$-$60$^{**}$ & $-$69$^{(3)}$ & 8/8 & 19/31 & 89/32 & 83/40\\ % 1st region: spots below 5km/s, 2nd region: spots above 10km/s
 G81.871 & $-$61$\pm$13 & $-$4 & $-$7 & $+$43$^{(4)}$ & 3 & 76 & 57 & 54\\
 G108.766 & $+$22$\pm$9 & $-$68 & $+$4 & - & 72 & - & 90 & 18\\
\hline
\end{tabular}
\tablefoot{Less reliable values are marked by italics: methanol maser emission in G48 consists of only two spots; therefore, the determined PA is not precise. $^*$Ex-OH masers in G43 form two diverse regions: eastern (blueshifted) and northwestern (redshifted), respectively. $^{**}$Ex-OH and methanol masers in G69 form two regions: northern blueshifted and southern redshifted emission. 
%$^{***}$ Methanol maser in G48 consists of only two spots, the determined PA is less reliable. JEST w caption
PA$_{\mathrm{out}}$ is estimated based on $^{(1)}$\cite{nagayama2015}, $^{(2)}$\cite{goddi2020}, $^{(3)}$\cite{kumar2004}, $^{(4)}$\cite{torrelles1997}.}
\label{tab:papa}
\end{table*}

\subsubsection{Zeeman triplets}\label{sec:triplets}
\citet{green2015} found 18 Zeeman triplet candidates among 112 Zeeman patterns in the ex-OH transition (detection rate of 16\%) when using ATCA with a spectral resolution of 24\,m\,s$^{-1}$. As we reported above, we detected 37 Zeeman pairs. The circularly polarized features (RHCP and LHCP) are the $\sigma$ components of the Zeeman splittings; they are shifted in the velocities by the magnetic field. A third component, the $\pi$ component, is the linearly polarized feature, unshifted in velocity.
When comparing the Zeeman pairs and the linearly polarized spots that we identified (Tables \href{https://doi.org/10.5281/zenodo.14865259}{C.1} and \href{https://doi.org/10.5281/zenodo.14865259}{C.2}), we find that in eight cases (21\%), namely Z$_3$ and Z$_4$ of G48, Z$_6$ and Z$_7$ of G49, Z$_1$, Z$_2$, and Z$_5$ of G69, and Z$_2$ of G81, we can suspect the existence of a Zeeman triplet; we mark these candidates with “T” in Table\,\href{https://doi.org/10.5281/zenodo.14865259}{C.2}. However, better angular and spectral resolution is needed to confirm that. 

In G25, the linearly polarized emission at the LSR velocity of 39.6\kms ~differs by only one spectral channel (0.1\kms) from the demagnetized velocity of 39.506\kms. A similar ambiguity is in Z$_1$ of G81. We mark such cases with “v” (for verification needed) in Table\,\href{https://doi.org/10.5281/zenodo.14865259}{C.2}, and again, observations with better angular and spectral resolution and good sensitivity are needed to verify the existence of Zeeman triplets. 

In G43, the linearly polarized spots coincide with the RHCP feature of Z$_2$, LHCP of Z$_8$, and interestingly with both circularly polarized features of Z$_3$ with an avoidance of the central velocities. Similarly, the avoidance of unshifted velocities is seen in G49 and Z$_4$, Z$_5$. That can indicate that this emission is associated with the $\sigma$ components, but also that the line-of-sight velocity can change due to effects % of some complications, 
such as turbulence.

\citet{green2015} showed that the flux densities of the $\pi$ components of the triplets are from 3\% to 85\% of the average circular component flux, with a median of 16\%, indicating the detection of $\pi$ components only toward the brightest Zeeman splittings. We do not confirm such a trend among our candidates for triplets. However, as we noted above, more sensitive and better angular and spectral resolution observations are needed to verify the existence of the potential Zeeman triplets. The lack of $\pi$ components may be related to the magnetic field orientation; if the magnetic field is parallel to the line of sight, the $\pi$ component is not seen.

%Surcis et al. 2015: seven HMYSOs, 19\% of 176 masing cloudlets show $m_{\mathrm{l}}$. The linear pol. vectors are well ordere excluding G174 without any polarization. The magnetic field of scales 10-100~au is preferentially oriented along the outflow axis.

%Surcis et al. 2019: seven HMYSOs, 47 out of 219 masing cloudlets show $m_{\mathrm{l}}$, 2 shows $m_{\mathrm{c}}$. Well-ordered linear polarized vectors. We confirm the magnetic field of scales 10-100~au is preferentially oriented along the outflow axis. Two cases where |B$_{||}$|>61~mG and >21 mG.

\subsection{Coincidence and avoidance of ex-OH and methanol masers}
\label{sec:coin}
\begin{figure*}
    \centering
    \includegraphics[width=\textwidth,trim={0 11.5cm 0 13cm},clip]{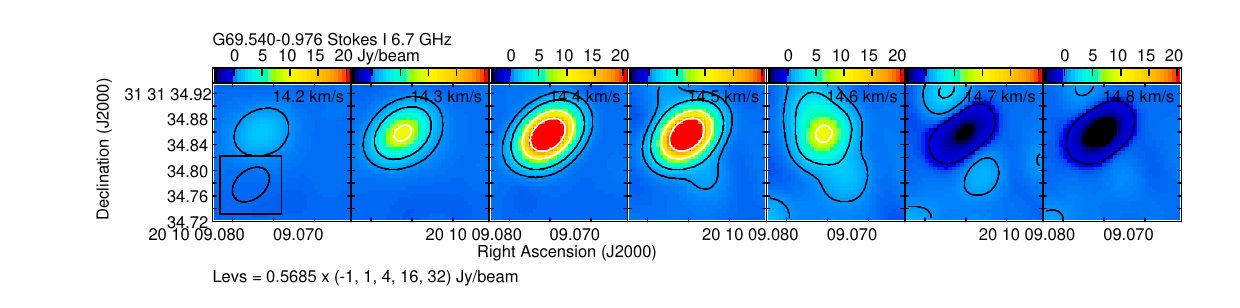}
    \includegraphics[width=\textwidth,trim={0 11.5cm 0 13cm},clip]{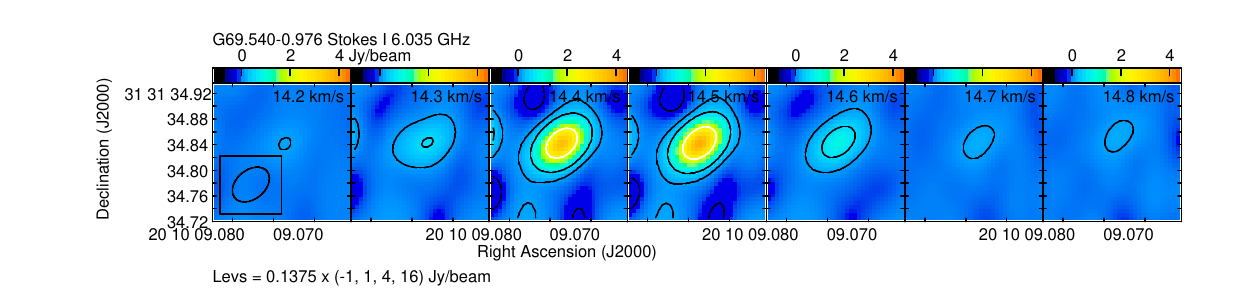}
    \caption{Channel images showing evidence for the coexistence of 6.7 GHz methanol (top) and 6.035~GHz ex-OH (bottom) masers obtained for G69. The ex-OH emission corresponds to Zeeman pair Z$_5$ as in Table\,\href{https://doi.org/10.5281/zenodo.14865259}{C.1}. Contours represent the emission; the first contours correspond to $3\sigma_\mathrm{rms}$ of the image, and the next contours are at 4, 8, 16, 32$\times3\sigma_\mathrm{rms}$. The first negative contour is also shown. The numbers in the top right corner of each panel correspond to the LSR velocities of each spectral channel.}
    \label{fig:coincidence_G69}
\end{figure*}

We note that the 6.7~GHz methanol maser emission is brighter than the ex-OH line in all targets, but the ratio $S_{\mathrm{p}}^{6.7}$/$S_{\mathrm{p}}^{6.035}$ is diverse, from 2.5 to 115 with a median of 15. In six of the targets, the 6.7~GHz methanol maser is spread over larger regions in the sky than the ex-OH maser, again with a broad range of size ratios, from 1.1 to 2700, with a median of 6.
In G43, G48, and G69, the ex-OH line emission has more complex structures and covers larger regions than methanol masers and in G108 the size of both transitions is similar. 
It is possible that the behavior stems from a lower kinetic temperature. The estimates of $T_{\mathrm{rot}}$ from CH$_3$CN lines give much lower temperatures for G43 and G69 than for G20 and G25, where methanol masers are more complex (see Appendix~\ref{sec:methylcynide}).

We carefully examined the data, searching for coexistence, according to our method described in the last paragraph of Sect.~\ref{sect:obser}, of both maser lines. In six targets, namely G20, G24, G34, G43, G69, and G81, we find coincidences of both transitions, considering the astrometric positions and overlap of the LSR velocities (see Fig.~\ref{fig:coincidence_G69} as the example). In total, in these six sources, we find 12 groups of masers that are coexisting via the same gas volume. In seven cases among these 12 groups, the ex-OH shows Zeeman splitting. 
In more detail, in G20, G69, and G81, we find cases in which a few ex-OH and methanol spots coincide, but other regions where they are separated. In these targets, the brightest methanol spots are not associated with ex-OH emission. In G24 and G34, all ex-OH spots coexist with methanol emission, and the ex-OH masers are located in compact areas. The strongest methanol masers coexist with ex-OH masers, with spot positions within 20~mas. The two maser transitions avoid each other in the four remaining targets; that is, G25, G48, G49, and G108. 

We checked correlations with other properties between areas showing the coexistence of both transitions and ones without this phenomenon. For this purpose, in all sources, we coupled the closest 6.7- and 6.035\ GHz cloudlets: 12 pairs with coexistence (as was noted above) and 8 couples with avoidance. Next, we checked the following parameters for all 20 pairs: the separations of the barycenters, the velocities of the peaks, and the ratios of brightness temperatures. We note that the coincidence of both maser transitions appears for the cloudlets with their barycenters separated by less than 205\,au and separations of their peak velocities below 0.7\kms.
We did not find any correlations with brightness temperature or the ratio of maximum brightness temperatures.

Based on the models by \citet{cragg2002}, the coincidence of both transitions happens when $n\rm{_{H_2}}$ is from 10$^{5}$ to 10$^{8.3}$\,cm$^{-3}$, $T\rm{_K}$ is below 70\,K, and $T\rm{_D}$ is larger than 100\,K. The number densities can be calculated from the $B_\mathrm{los}$ estimated from the 37 Zeeman pairs identified (see Sect.~\ref{sec:polarizationZP} and  Table\,\href{https://doi.org/10.5281/zenodo.14865259}{C.1}). In the case of four pairs (two in G108, one in G49, and one in G81), $n_\mathrm{H_2}$ exceeds a value of 10$^{8.3}$\,cm$^{-3}$. In fact, in all these cases, no coexistence appears for the maser lines. For the other 33 Zeeman pairs, avoidance and coincidence must be related to gas and/or dust temperatures.

In sources G20, G43, G69, and G81, there are regions of coincidence and also regions where ex-OH emission occurs without methanol transition. We have already noted that the redshifted feature (Z$_3$) in G81 does not have a methanol counterpart because of the high density of the gas. In the other three sources, we know the kinetic temperature has to be below 70~K from the model. Thus, the reason for the presence of ex-OH without 6.7\,GHz masers could be a decrease in the dust temperature below 100~K on scales of hundreds of astronomical units (around~500\,au for G20, 1000~au for G43, and 400\,au for G69).

In sources G20, G24, G34, G69, and G81, there are regions of coincidence and regions where methanol emission occurs without ex-OH transition. We know T$\rm{_D}$ has to be above 100~K. Thus, a possible reason for the absence of the ex-OH maser could be an increase in the kinetic temperature above 70\,K, whereby the condition $T\rm{_K} < T\rm{_D}$ has to be conserved. The scale of variability of $T\rm{_K}$ is around~170~au for G20 and G34, 1000~au for G69, and 450~au for G81. 

As has been noted, in G25, G48, G49, and G108 we do not find any coexistence of the maser transitions. For G108, as for G81, we inferred a number of high densities in the volume of gas where ex-OH occurs. G25 and G49 have very bright methanol maser emission, which requires a strong IR source. From Table\,\href{https://doi.org/10.5281/zenodo.14865259}{C.3}, it can be seen that the G25 is the most massive YSO and has the highest bolometric luminosity. Hence, the main factor influencing the avoidance of both transitions in these two sources may be the distance from the YSO. 
In regions that are closer to the central object, and thus hotter, the 6.7\,GHz maser occurs, and in distant and cooler regions, the 6.035~GHz maser transition appears. In G48, the brightness of both transitions is similar, but we observe a significant shift in the velocity domain at~1.9\kms ~between methanol and ex-OH. This supports the hypothesis that these transitions appear in different volumes of gas. This target has the weakest bolometric luminosity and the smallest mass of YSO, which explains why a very weak methanol maser appeared, consisting of only two spots (Table\,\href{https://doi.org/10.5281/zenodo.14865259}{C.3}).

We find that the distributions of both masers are roughly parallel in seven targets (G24, G25, G34, G43, G49, G69, and G81), |PA$_{\mathrm{6.7}}$-PA$_{\mathrm{6.035}}$| < 31\degr (Table\,\ref{tab:papa}). Kolmogorov-Smirnov tests for these two parameters confirm a nonrandom relationship between PA$_{\mathrm{6.7}}$ and PA$_{\mathrm{6.035}}$ values. This indicates that both masers could be associated with the same kinematic structures, although they do not necessarily trace the same parts.
The blue- and redshifted maser velocities are consistent with the velocities of SO, CH$_3$CN, and HC$_3$N lines in G25, G43, and G69. That may imply that the masers arise from a disk but further data including proper motion studies are needed to verify this scenario. In Sect. \ref{sec:introd}, we pointed out that methanol masers in G69 were also considered to be associated with the outflow \citep{sugiyama2011}.
Among the other sources, we did not find signs of rotation, based on ALMA data presented in the Sect.\,\ref{sec:couterparts}. The masers in G24 have similar PA$_{6.7}$ and PA$_{6.035}$, perpendicular to the known outflow, which may indicate a disk seen edge-on. For G34, we see a ring-like maser morphology, but the lack of other data prevents us from determining the kinematic structure. The situation is similarly unresolved for G49 and G81. 
Considering the findings of the counterpart star formation tracers presented in Sect.~\ref{sec:couterparts}, we notice that circular cores of millimeter dust continuum emission overlaps maser emission in G20, G24, G43, and G69. All these sources show the coincidence of both maser transitions. G43 and G69, which show more complex ex-OH emission, have much brighter 1.3\,mm cores, of 159\,mJy and 236\,mJy, respectively, than G20 and G24, which have a brightness of 20\,mJy and 3\,mJy, respectively. In G49 and G81, maser emission is clearly separated from irregular millimeter cores. Those two sources are the brightest in our sample, but 6.7 and 6.035\,GHz masers coexist only in G81.
We cannot determine any association between kinematic structure and/or morphology and the coexistence of both maser transitions. In four sources (G24, G25, G43, and G69), masers seem to be associated with the disk, but in G25, we notice the avoidance of both maser transitions, and in the other three sources, we found partial coincidence.

Relating to our analysis of the evolutionary stages of our sample (Sect.~\ref{sec:evolve}), taking into account all three evolutionary indicators, the most evolved sources are sequentially G69, G43, and G48. All three characterize the most complex distributions of ex-OH compared to methanol masers. G69 has very bright 6.7\,GHz emission, $\sim$100\,Jy, but G43 and G48 are relatively weak. G69 and G43 show the coexistence of the two maser transitions, and G48 shows avoidance. We found signs of rotating disks for three sources in our sample. Among them, two are the most evolved: G69 and G43. We can also notice that, in our sample, only these three, the most evolved sources, have more complex and extended ex-OH maser structures than methanol maser structures.} 
The next sources in the evolutionary sequence are G20, G24, G25, and G81. Three of them show coexistence, and one, G25, shows avoidance. They also have various morphologies, and only for G25 do we find the sign of a rotating disk. The last three sources, G49, G34, and G108, are in an early evolutionary phase (our interpretation for G49 could be wrong because it is not clear with which core masers emissions are associated). These sources are extremely different. G49 is the brightest source in the 6.7\,GHz line in our sample and does not show a coincidence with the ex-OH line, similarly to G108, which has one of the weakest methanol maser emissions. Both sources also show a similarly complex morphology for both transitions. On the other hand, G34 shows very compact ex-OH emission, which does coincide with methanol emission.

To summarize, we do not notice a relationship between the phase of the evolution of YSO and the presence of regions with coexistence or avoidance of methanol and ex-OH masers. Some of the most evolved sources, like G43 and G69, show coincidence, but the next most evolved one, G48, shows avoidance. Also, two sources in the very early evolutionary phase, G34 and G108, show coincidence and avoidance, respectively. The factors that impact the appearance of the two transitions seem to be due to local changes in gas and dust temperatures and gas density, on the scales of a few hundred astronomical units, correlated with the distance between the maser and star and the brightness of the YSO. 

\section{Conclusions}
We examined ten HMYSOs in detail, considering two maser transitions: 6.7\,GHz methanol and 6.035\,GHz OH. The ex-OH maser line was imaged for the first time for eight of them. 
The 6.7\,GHz methanol maser emission is brighter than the ex-OH line in all targets and is spread over larger regions in the sky than ex-OH masers in six of these. 
In three HMYSOs (G43, G48, and G69), the ex-OH masers have more complex structures and cover larger regions than methanol masers, and in one (G108) the size of both transitions is similar.

We identified regions where the two maser transitions coincided or showed avoidance. Comparing our result with archival ALMA data, we deduced that avoidance happens due to local changes in temperatures and/or densities, which are directly correlated with the brightness and distance of the masers from the YSO and not with the evolutionary stage of the HMYSO. We also attempted to identify kinematic structures, but we did not notice a correlation with the coexistence of both masers.

We detected Zeeman splitting of the ex-OH transition in eight HMYSOs, leading to magnetic field estimates of a few milligauss. In two cases, we observed a reversal of magnetic field lines along the line of sight. We detected linearly polarized ex-OH emission in nine HMYSOs and were able to compare orientations of the magnetic fields with the directions of outflows on the sky-plane, but the results are not coherent enough. We detected eight possible Zeeman triplets with two circularly polarized features at velocities bracketing a linearly polarized feature; however, observations with better angular and spectral resolution are needed to verify this result.

\section*{Data availability}
An extra appendices containing tables of polarization details, figures of masers distributions, and counterparts descriptions are available on Zenodo at: \url{https://doi.org/10.5281/zenodo.14865259}.

\begin{acknowledgements}
We acknowledge support from the National Science Centre, Poland through grant 2021/43/B/ST9/02008. We thank a referee for the very detailed and constructive revision of the manuscript. 
e-MERLIN is a National Facility operated by the University of Manchester at Jodrell Bank Observatory on behalf of STFC
This paper makes use of the following ALMA data: ADS/JAO.ALMA2021.1.00311.S, ADS/JAO.ALMA2019.1.00059.S., and ADS/JAO.ALMA2015.1.01596.S. ALMA is a partnership of ESO (representing its member states), NSF (USA) and NINS (Japan), together with NRC (Canada), NSTC and ASIAA (Taiwan), and KASI (Republic of Korea), in cooperation with the Republic of Chile. The Joint ALMA Observatory is operated by ESO, AUI/NRAO and NAOJ.
The National Radio Astronomy Observatory is a facility of the National Science Foundation operated under cooperative agreement by Associated Universities, Inc.
This work is based on observations made with the \textit{Spitzer} Space Telescope, which is operated by the Jet Propulsion Laboratory, California Institute of Technology under a contract with NASA.
This research has made use of the NASA/IPAC Infrared Science Archive, which is funded by the National Aeronautics and Space Administration and operated by the California Institute of Technology.
Herschel is an ESA space observatory with science instruments provided by European-led Principal Investigator consortia and with important participation from NASA.

\end{acknowledgements}

\bibliography{librarian}{}
\bibliographystyle{aa}
 
%\Online
\begin{appendix}

\section{Methyl cyanide}
\label{sec:methylcynide}
Methyl cyanide ($\mathrm{CH_3CN}$) $J=12-11$ rotational transition could be analyzed for a number of sources using the data from the ALMA Science Archive. Under the assumption that the lines are optically thin, a rotational temperature can be derived from the line intensities of various $K$ components. If the lines are optically thick, the various $K$ intensities versus the upper energy level will deviate from a linear slope, and the slope may flatten, which leads to overestimated temperatures \citep{goldsmith1999}. When also the $\mathrm{^{13}C}$ isotopolog, $\mathrm{CH_3^{13}CN}$, is detected, the optical depth of $\mathrm{CH_3^{12}CN}$ can be derived from the $K$ line intensity ratio under the assumptions that $\mathrm{CH_3^{13}CN}$ is optically thin and the $\mathrm{^{12}C/^{13}C}$ line ratio of 50 \citep{wilson1994}. See \cite{jimenez2017} for a detailed explanation.  

Following the analysis outlined in \cite{pankonin2001}, \cite{araya2005} and \cite{rosero2013} we analyzed available ALMA archival data. We used eight Gaussians to fit contemporaneously the $K=0$ to $K=7$ components, constraining the line width of all lines to be the same as for the bright and unblended $K=3$ component, and fixing positions of the lines to the rest frequencies, after taking into account the LSR velocity of the target. The results of the fits are given in Table~\ref{tab:ch3cn}.

Strong deviations from linear were found for G20 and G25, less so for G43, and least for G69. Since also the $K=2$ $\mathrm{CH_3^{13}CN}$ could be detected, the optical depth of the main isotopolog ($\tau_{^{12}C}$) could be calculated. We find that $\mathrm{CH_3^{12}CN}$ is indeed optically thick for all our sources, more so for those with a stronger deviation from linear. Therefore, all of our estimated temperatures are overestimated, and our column densities are underestimated. On the argument of beam dilution, which is often invoked in the past single-dish analysis of $\mathrm{CH_3CN}$, we are less worried because we use ALMA interferometric data. 
We measured the spectra in ellipses chosen to include most of the $\mathrm{CH_3CN}$ emitting core which overlapped or was closest to the maser emitting region. The smallest minor axis was 0.4 arcsec and the largest major axis was 1.2 arcsec, these being (slightly) larger than the beam sizes. The emission of the $\mathrm{CH_3CN}$ $J=12-11$ transition observed by ALMA is thus slightly extended, making it plausible to use a beam filling factor of unity.

%\begin{table*}[h!]
\begin{sidewaystable}
\caption{\label{tab:ch3cn} Derived parameters of the $\mathrm{CH_3CN}$ $J=12-11$ ALMA observations.}
\centering
\begin{tabular}{cccccccccccccccc}
\hline
Source & vlsr & linewidth & \multicolumn{8}{c}{$\int T \delta v$} & Trot & N &$\mathrm{^{13}C}$ & ratio & $\tau_{^{12}C}$ \\
       &       &          &$K=0$ & $K=1$ & $K=2$ & $K=3$ & $K=4$ & $K=5$ & $K=6$ & $K=7$ & \\
& (km s$^{-1}$) & (km s$^{-1}$) & \multicolumn{8}{c}{(K km s$^{-1}$)} & K & (e$^{14}$ cm$^{-2}$) & & & \\
\hline
G20 &72.5 &3.4 &87.5 &80.2 &87.9& 91.2  &55.4 &50.6 &43.2 &21.9 &335(79) &25.1(3.8)& y & 7.7 &6.5 \\% 0.22x0.20 semi axis, at 18:27:44.5656168117, -11:14:54.1387759072, ratio 12/13=10.8300383/1.39762996
G25& 42.0 &4.2 &456.9 & 491.1& 514.6 &566.4  & 407.0 &345.9 &362.6 &150.4 &446(143) &229(32) &y &7.6  &6.5\\% 12/13 ratio=48.81183686/6.38892872, ellipse 0.61x0.46  18:34:20.9138785797, -5:59:42.2574205802
G43 &12.0& 2.1 &37.4 &30.4 &28.9 &31.6 &14.8  &9.3  &9.5  &3.4 &177(15) &4.4(0.5) &y &8.4 &5.9 \\%0.40x0.386 at 19:10:11.0624820788, 9:05:20.3217580885 13/12 ratio = 5.52056603/0.65674296
G69& 14.0 &1.7 &167.3 &155.2 &145.1 &162.2  &95.1  &64.7  &56.8  &14.5 &180(18) &23.7(2.9) &y &14.8 &3.4 \\% 0.37x0.387 asec at 20:10:09.049499, 31:31:35.041215 13/12 ratio = 33.88188234/2.2777968
\hline
\end{tabular}
%\addtocounter{table}{-1}
%\end{table*}    
\end{sidewaystable}

\end{appendix}

\end{document}